\documentclass[fleqn,usenatbib]{mnras}

\usepackage[T1]{fontenc}
\usepackage{ae,aecompl}
\usepackage{xtab}
\usepackage{courier}
\usepackage{multirow}
\usepackage{times}
\usepackage{upgreek}
\usepackage{graphicx}
\usepackage{epstopdf}
\usepackage{amssymb}
\usepackage{amsmath}
\usepackage{ulem}
\usepackage[section]{placeins}

\usepackage{mathptmx}
\usepackage{txfonts}

\usepackage[T1]{fontenc}
\usepackage{ae,aecompl}

\usepackage[table]{xcolor}
\usepackage{tabulary}

\usepackage{dirtytalk}
\usepackage{comment}

\newcommand{\uHz}{$\mathrm{\mu}$Hz}
\newcommand{\tess}{{\it TESS}}
\newcommand{\kep}{{\it Kepler}}
\newcommand{\gaia}{{\it Gaia}}
\newcommand{\hwvir}{{HW\,Vir}}
 
\newcommand{\cmss}{\,\mbox{$\mbox{cm}\,\mbox{s}^{-2}$}}    

\definecolor{light1}{gray}{0.75}
\definecolor{light2}{gray}{0.85}
\definecolor{light3}{gray}{0.95}

\title[\tess\ sdBVs in the northern ecliptic hemisphere]{A search for variable subdwarf B stars in \tess\ Full Frame Images\\
II. Variable objects in the northern ecliptic hemisphere}

\author[A.S. Baran et al.]{
A.S.\,Baran,$^{1,2,3}$\thanks{E-mail: andrzej.baran@up.krakow.pl}
S.K.\,Sahoo,$^{1,4}$
S.\,Sanjayan$^{1,4}$
and J.\,Ostrowski$^{1}$
\\
$^{1}$ARDASTELLA Research Group, Institute of Physics, Pedagogical University of Krakow, ul. Podchor\c{a}\.zych 2, 30-084 Krak\'ow, Poland\\
$^{2}$Department of Physics, Astronomy, and Materials Science, Missouri State University, Springfield, MO 65897, USA\\
$^{3}$Embry-Riddle Aeronautical University, Department of Physical Science, Daytona Beach, FL\,32114, USA\\
$^{4}$Nicolaus Copernicus Astronomical Centre of the Polish Academy of Sciences, ul. Bartycka 18, 00-716 Warsaw, Poland\\
}

\date{Accepted XXX. Received YYY; in original form ZZZ}

\pubyear{2021}

\begin{document}
\label{firstpage}
\pagerange{\pageref{firstpage}--\pageref{lastpage}}
\maketitle

\begin{abstract}
We report the results of our search for pulsating subdwarf B stars in Full Frame Images collected during Year 2 of the \tess\ mission and covering the northern ecliptic hemisphere. This is a continuation of our effort we presented in Paper I. We found 13 likely new pulsating subdwarf B stars, 10 pulsating candidates that are identified as other hot subdwarfs, and 30 spectroscopically unclassified objects that show amplitude spectra typical of pulsating subdwarf B stars. We found 506 variable objects, most of them spectroscopically unclassified, hence their specific variability class yet to be confirmed. Eclipsing binaries with sharp eclipses sample comprises 33 systems. For 12 of them we derived precise orbital periods and checked their stabilities. We identified one known and five new candidate \hwvir\ systems. The amplitude spectra of the 13 likely sdB pulsators are not rich in modes, hence any further analysis is not possible. However, we selected three candidates for pulsating subdwarf B stars that show the richest amplitude spectra and we performed a mode identification deriving modal degrees of most of the detected modes. In total, in both ecliptic hemispheres, we found 15 likely pulsating pulsating subdwarf B stars, additional 10 candidates for pulsating subdwarf B stars, 66 other variable subdwarf B stars, 2076 spectroscopically unconfirmed variable stars, and 123 variable non-sdB stars. 
\end{abstract}


\section{Introduction}
\citet[][hereafter: Paper I]{sahoo20} reported their findings in the southern ecliptic hemisphere f 1807 variable stars, including two likely new pulsating subdwarf B stars (sdBV), 26 non-pulsating sdBs, and 83 spectroscopically unclassified objects whose amplitude spectra show peaks in the sdBV g-mode region. The bulk of the variables have no spectral classification so the result of \citet{sahoo20} defines a catalog of objects that require spectroscopic verification.

We continued our effort to search for variable subdwarf B stars (sdB) in the northern ecliptic hemisphere observed by \tess\ and stored in Full Frame Images (FFI). Our main goal is to complete the census of variable sdBs in the \tess\ field-of-view. Many targets have been incorporated in the short cadence, either 2\,min or 20\,sec, monitoring, however this observing mode is limited in number of targets and therefore not all sdBV candidates will be observed. Our search is not subject to any pre-approval since we use FFI images that are stored automatically and all targets in those images are available for data processing. Handling all objects detectable in the FFIs, although desirable, is technically difficult (processing would take months if not years), hence we focus on targets listed in the sdB catalogue described by \citet{geier20}.

Subdwarf B\,(sdB) stars are identified as objects located at the blue end of the horizontal branch, often called the extreme horizontal branch (EHB), in the Hertzsprung-Russell diagram \citep{heber16}. These stars are compact in size, with surface gravities, $\log$ (g/\cmss), of 5.0 to 5.8, which translates into radii of 0.15\,--\,0.35\,R$_{\sun}$. SdBs are blue due to their high effective surface temperature\,(T$_{\rm eff}$) ranging between 20,000 and 40,000\,K. The sdB stars have masses 0.47\,M$_{\sun}$ on average, which is sometimes called the canonical mass \citep{heber16}.

Pulsations in sdBs were discovered observationally by \citet{kilkenny97} and theoretically by \citet{charpinet97}. The pulsations were found at both low and high frequencies. The low frequencies (long periods of hours) are explained by gravity modes, while the high frequencies (short periods of minutes) are explained by pressure modes \citep{fontaine03}. To increase a sample of known pulsating sdB (sdBV) stars, it was essential to make an effort at more discoveries. First discoveries were made from the ground and detections were limited to sdBVs showing pressure modes. They were easier to detect because of their higher observable amplitudes and shorter periods as compared to sdBVs showing gravity modes. Only a handful of sdBVs were found to be pulsating in both types of modes, with Balloon\,090100001 being the best example \citep{baran09}. To date, there are around 130 sdBV found in both ground and \kep\ space data \citep{holdsworth17,reed18b}. Additional 200+ sdBVs are detected in \tess\ short cadence data and the full list will be reported elsewhere.

SdBVs are found in both open \citep{reed12} and globular \citep{randall09} clusters but mostly in the Galactic field. We still lack a complete all-sky search for sdBVs, which prevents a comprehensive analysis of pulsation properties correlated with stellar population to conclude the location of instability strip(s) or physical parameters (especially masses). An all-sky search for sdBVs, along with \gaia\ parallaxes would tell us about the distribution of these stars in the Galaxy and contribute toward pulsation-stellar population relationships.

We aimed at selecting sdBs and sdB candidates that were not included in the predefined target list, producing time-series data directly from FFIs and making mode identifications of the most suitable cases showing pulsations. The results we present here are based on data collected during Year 2 in Sectors 14--26. Each sector is monitored for about 27\,days. In Section\,\ref{sec:gaia} we describe the source of our targets, the selection process and data processing. In Section\,\ref{sec:results} we present objects with found variables. Section\,\ref{sec:modeID} reports our mode identification effort, followed by Section\,\ref{sec:summary} that summarizes our results.

\section{Target selection and data processing}
\label{sec:gaia}
To select sdB candidates we used the sdB database reported by \citet{geier20}, which was prepared based on ESA \gaia\ Data Release 2 (DR2) and several ground-based multi-band photometry surveys. \citet{geier20} used color indices, absolute magnitudes and reduced proper motions to select the most suitable sdB candidates. The database is limited to \gaia\ G\,mag\,=\,19 and contains 39\,800 objects. From this sample we selected 15\,191 objects located in the northern ecliptic hemisphere and the \tess\ field of view. Using \gaia\ IDs and target coordinates we applied TOPCAT \citep{taylor05} and rejected 548 targets that were assigned to be observed in the short cadence mode. Then, we used {\it Tesscut} \citep{brasseur19} to collect sector information targets in our sample that will be observed in, and targets with no sector assignment were also rejected. Finally, we filtered targets observed in sectors 14-26 and we ended up with 5\,816 targets. It turned out that 464 targets have no useful data, so these were also rejected from our sample, hence the final number of targets was 4\,804. For completeness, we have included targets with non-sdB spectral classification as it may happen that some of these already classified objects will be reclassified as sdBs with further analysis. In addition, even if the stars will not be reclassified as sdBs, other researchers may find it useful to have these objects identified as variables.

The extraction process is exactly the same as described by \citet{sahoo20}. First, we used the {\it Eleanor} \citep{feinstein19} for downloading and processing data directly from \tess\ FFIs, which we downloaded from the "Barbara A. Mikulski Archive for Space Telescopes" (MAST). We specified a square target mask of 15\,pixels on side and a square background mask of 31\,pixels on side. {\it Eleanor} delivers raw and corrected time-series data. The raw data is a sky-subtracted simple aperture photometry, which is basically a sum of all flux within an optimal aperture for each timestamp. The corrected data account for known satellite artifacts. We extracted both data sets and have chosen the one that shows better signal-to-noise (S/N).

Then, we used {\it lightkurve} python package \citep{hedges18} to detrend and remove outliers from time-series data. Finally, we normalized fluxes by calculating (f/$\bar{\rm f}-$1)$\cdot$1000, where f and $\bar{\rm f}$ stands for a flux and median flux respectively, reporting amplitudes in {\it parts per thousand} (ppt). We remind the reader that the {\it Eleanor} does not account for a crowding metric correction and a flux fraction, which often results in an overestimated average flux and a diluted amplitude of a flux variation. Therefore, the amplitudes are often not realistic. In case of sdBV stars, the amplitudes are not so important, but they are for accurate modelling of eclipsing binaries or classical pulsators.

We searched for a flux variation in amplitude spectra we calculated for each target. We used a threshold of S/N ratio of 4.5 \citep{baran15}. The cadence of 30\,min defines the Nyquist frequency at 277\,${\mu}$Hz (24\,c/d). This limits our search to gravity mode sdBVs only.

\section{The zoo of our findings}
\label{sec:results}
We detected significant flux variations in 506 objects, including 53 sdBs, 33 subdwarfs (sd), 374 not spectroscopically classified, and 46 classified as non-sdBs. To identify sdB objects we used the sdB database \citep{geier20} and the Simbad database \citep{wenger2000}. We listed all variable objects in  Tables\,\ref{tab:sdB}--\ref{tab:nospecft}, including those in on-line material, which provide basic information on our findings and possible contamination. The large square pixels, 21\,arcsec on side, cause serious issues in crowded regions of the sky as an optimal aperture, typically 2--3 pixels, may contain neighboring objects. In case the optimal aperture covers  neighboring sources, and we are not positive about which object shows a flux variation, we made remarks according to the following rule. If an optimal aperture contains more than five objects we marked it as \say{crowded region}, if the optimal aperture is densely covered by stars, {\it e.g.} a cluster region, we marked it as \say{very crowded region}, if a few objects\,(<\,5) were spotted, we specified the number of objects within a given radius. In case of just one neighboring object within an optimal aperture, we provided the distance to the object, and we provided its designation, if any. If we found that any of our targets is already known as a variable star, we provided a reference. Even though these targets are not new variables, we included them for completeness.

\subsection{Spectroscopically confirmed sdB variables}
\label{sec:sdB}
\begin{figure*}
\includegraphics[width=1\textwidth]{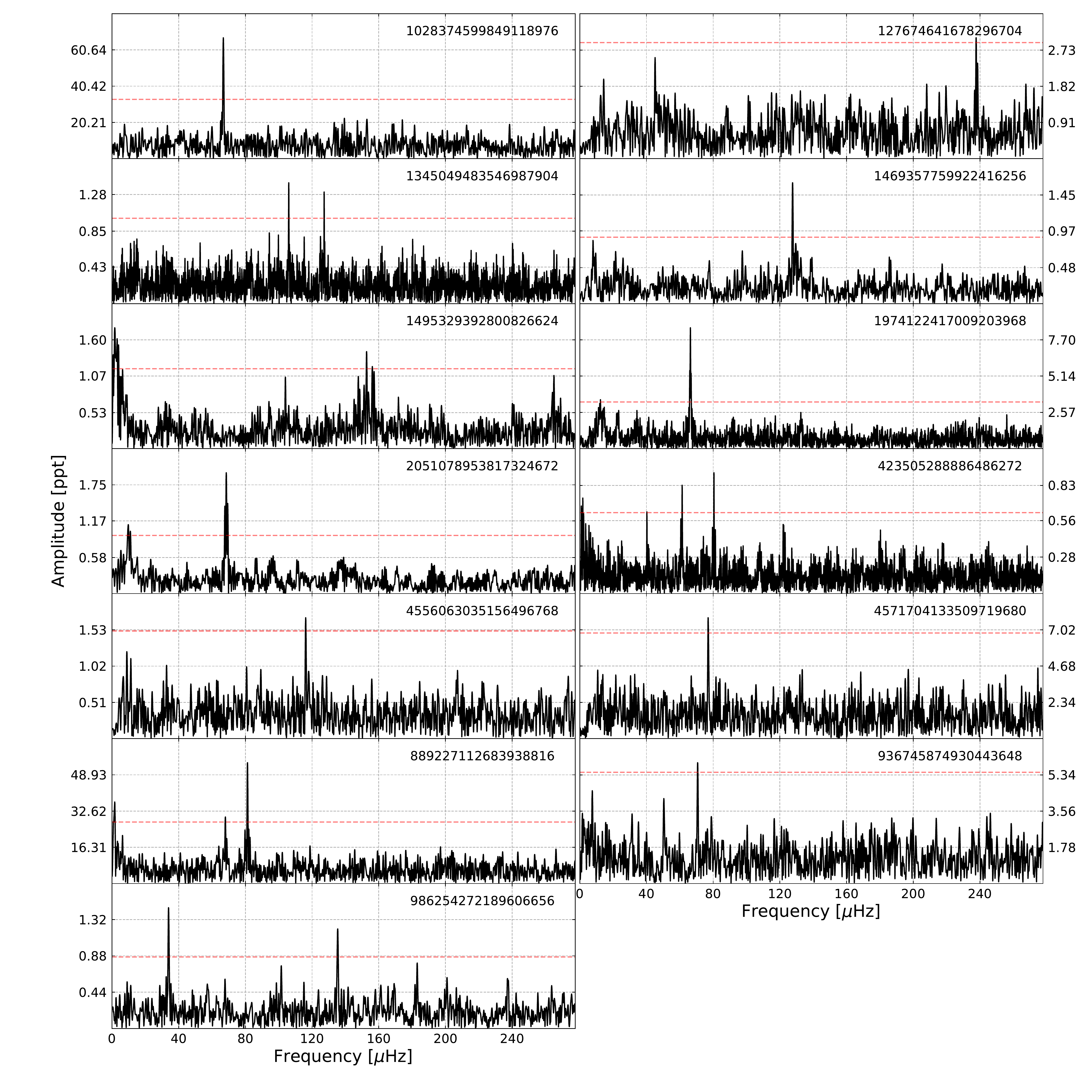}
\caption{Amplitude spectra of 13 sdBV stars listed in the first group of Table\,\ref{tab:sdB}. Horizontal red dashed lines indicate S/N\,=\,4.5 detection threshold (all relevant figures). The long numbers in all figures denote \gaia\,DR2 numbers of specific stars.}
\label{fig:sdBV}
\end{figure*}

\begin{figure*}
\includegraphics[width=1\textwidth]{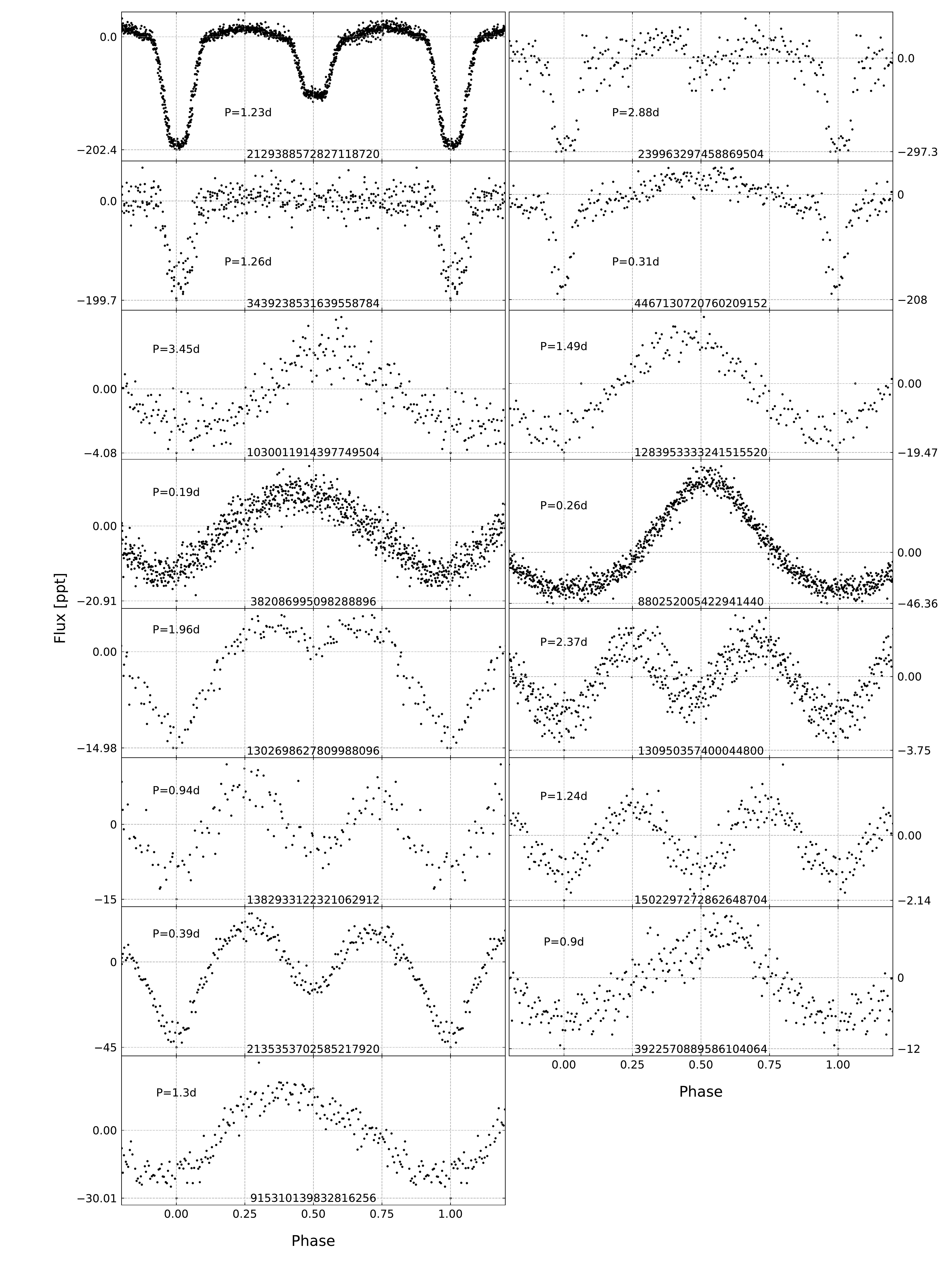}
\caption{Phased and binned time-series data of variable sdB stars listed in the second group of Table\,\ref{tab:sdB}.}
\label{fig:sdBp}
\end{figure*}

\begin{figure*}
\includegraphics[width=1\textwidth]{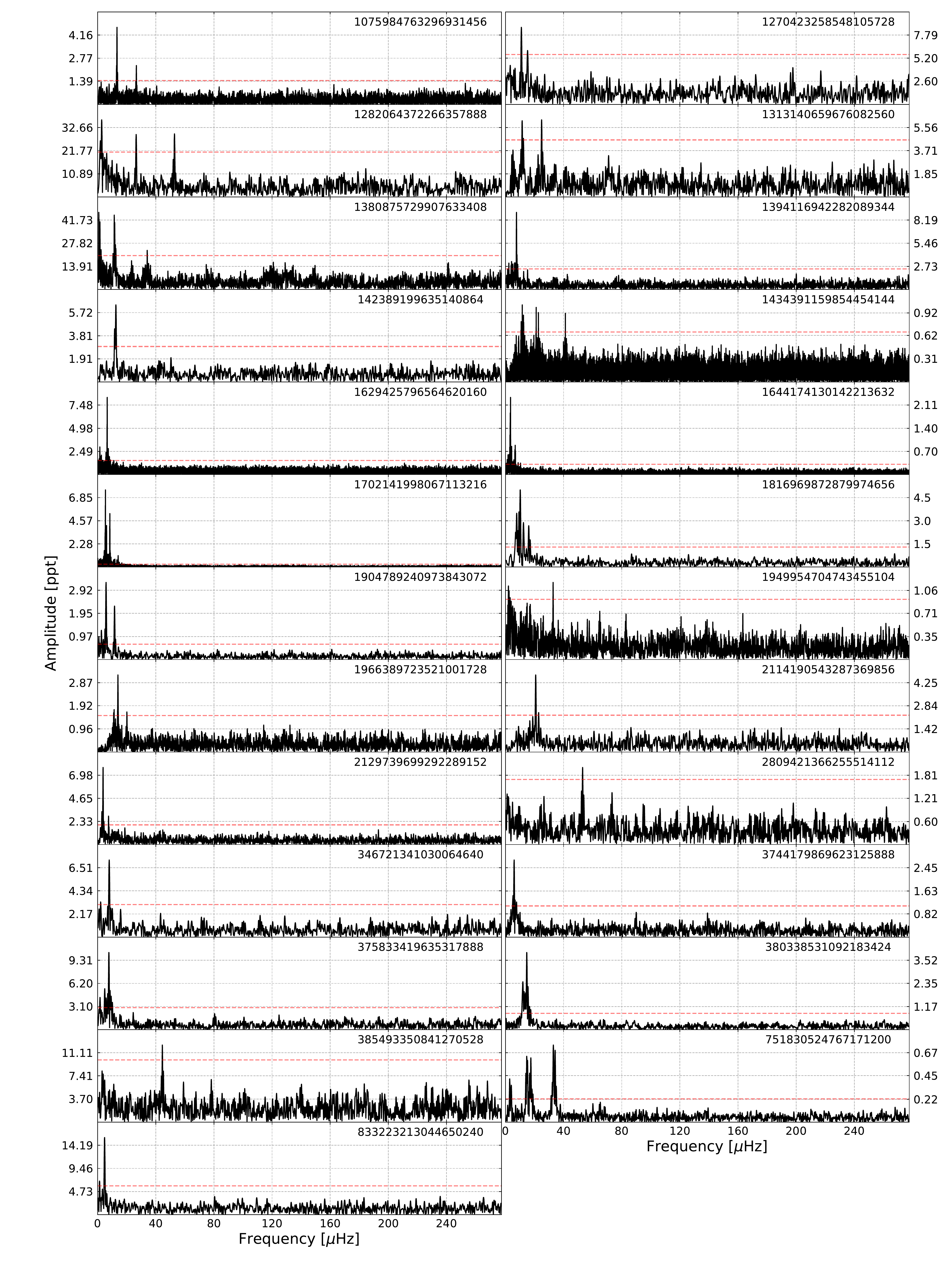}
\caption{Amplitude spectra of variable sdB stars listed in the third group of Table\,\ref{tab:sdB}.}
\label{fig:sdBf}
\end{figure*}

We found 53 variable objects classified as \say{sdBs} and we list them in Table\,\ref{tab:sdB}. This table is divided into three groups. The first group contains 13 objects with peaks above 60\,\uHz\ and we identified them as likely pulsating sdBs. Our interpretation may not be correct, since binary peaks can also be detected above 60\,\uHz\ and that is why we call this group "likely sdBV stars", as better quality data must confirm our interpretation. We show the amplitude spectra of these objects in Figure\,\ref{fig:sdBV}. All these objects show poor g-mode spectra and further analysis related to pulsations in sdBs cannot be performed. The second group contains 15 sdBs that show clear orbital flux variation and do not show any pulsation signature. We found four eclipsing binaries with sharp eclipses, four objects showing one maximum, which can be interpreted as {\it e.g.} a reflection effect, five objects showing two maxima, which can be either eclipsing and/or ellipsoidal systems, and two systems showing one asymmetric maximum that is typical of classical pulsators. We show the phased time-series data in Figure\,\ref{fig:sdBp}. \gaia\ DR2\,4467130720760209152 shows \hwvir\ flux variations, while \gaia\ DR2\,880252005422941440 was, the most likely, mistakenly classified as RR\,Lyrae star by \citet{drake14}. The optimal aperture for \gaia\ DR2\,382086995098288896 overlaps with \gaia\ DR2\,382086995093288256, which has been reported as variable by \citet{drake14}. The period we find is half the one cited by \citet{drake14} and most likely the variation we find comes from the latter object. The third group contains 25 sdBs that show flux variations that cannot be pulsations, while the phased time-series data either do not show significant variation or multiple peaks that make the phased time-series looks messy. We plot the amplitude spectra in Figure\,\ref{fig:sdBf}. Most of the time they show related peaks likely indicating a binary nature of the systems, while single peaks below 60\,\uHz\ are atypical for sdBVs. \gaia\ DR2\,2129739699292289152 is found to be variable by \citet{ostensen11}.

\begin{table*}
\centering
\caption{Basic information of 53 objects classified as sdBs. Time-series data or amplitude spectra of these objects are plotted in Figures\,\ref{fig:sdBV}, \ref{fig:sdBp} and \ref{fig:sdBf}. Objects in all tables and figures are sorted by \gaia\ ID, not accounting for the length of the numbers.}
\label{tab:sdB}
\rowcolors{1}{light3}{}
\resizebox{0.85\textwidth}{!}{\begin{tabular}{lrlcccl}
\hline
\rowcolor{light1}
& & & \multicolumn{1}{c}{G} & & \multicolumn{1}{c}{Period} & \\
\rowcolor{light1}
\multicolumn{1}{c}{\multirow{-2}{*}{\gaia\ DR2}} & \multicolumn{1}{c}{\multirow{-2}{*}{TIC}} & \multicolumn{1}{c}{\multirow{-2}{*}{Name}} & \multicolumn{1}{c}{[mag]} & \multicolumn{1}{c}{\multirow{-2}{*}{Sector}} & \multicolumn{1}{c}{[days]} &  \multicolumn{1}{c}{\multirow{-2}{*}{Remarks}}\\
\hline
\rowcolor{light2}
\multicolumn{7}{c}{sdBVs -- Figure\,\ref{fig:sdBV}}\\
\hline
1028374599849118976 & 802232206 & SDSS\,J082428.41+512601.6 & 18.7153 & 20 & 0.1734 & - \\
127674641678296704 & 353892824 & KUV\,02281+2730 & 15.1482 & 18 & 0.0487 & - \\
1345049483546987904 & 159850392 & GALEX\,J17566+4125 & 14.2792 & 25,26 & 0.08 - 0.13 & - \\
1469357759922416256 & 321423000 & SDSS\,J132432.37+320420.9 & 16.6387 & 23 & 0.0907 & - \\
1495329392800826624 & 23746001 & PG\,1350+372 & 14.3133 & 23 & 0.06 - 0.09 & - \\
1974122417009203968 & 305867900 & LAMOST\,J213911.57+453915.2 & 16.5692 & 15,16 & 0.1745 & 4 objects within 21" \\
2051078953817324672 & 122366140 & SDSS\,J192059.78+372220.0 & 15.7712 & 14 & 0.1689 & 4 objects within 21" \\
423505288886486272 & 445380420 & LAMOST\,J005821.83+553750.9 & 13.2004 & 17,18 & 0.12 - 0.35 & 2 bright objects within 21" \\
4556063035156496768 & 363766470 & HS\,1741+2133 & 14.0105 & 26 & 0.0996 & object 7" away \\
4571704133509719680 & 1309303943 & SDSS\,J170256.38+241757.9 & 19.0562 & 25 & 0.1504 & 5 objects within 21" \\
889227112683938816 & 741122759 & LAMOST\,J065538.32+312339.8 & 17.0121 & 20 & 0.1 - 0.2 & - \\
936745874930443648 & 741726760 & SDSS\,J075937.15+541022.2 & 17.7817 & 20 & 0.1638 & object 14" away \\
986254272189606656 & 742806233 & LAMOST\,J073935.74+564233.2 & 16.6757 & 20 & 0.08 - 0.4 & - \\
\hline
\rowcolor{light2}
\multicolumn{7}{c}{variable sdBs -- Figure\,\ref{fig:sdBp}}\\
\hline
2129388572827118720 & 1882909457 & Kepler\,J19211+4759 & 17.6946 & 14,15 & 1.2319 & - \\
239963297458869504 & 642677463 & SDSS\,J030749.25+411401.6 & 17.6208 & 18 & 2.8795 & - \\
3439238531639558784 & 172171754 & KUV\,06290+3235 & 16.2259 & 20 & 1.2607 & - \\
4467130720760209152 & 356085716 & PG\,1628+181 & 15.3912 & 25 & 0.3094 & \hwvir \\
1030011914397749504 & 802252743 & SDSS\,J084556.15+542357.6 & 18.8719 & 20 & 3.4529 & bright object 32" away \\
1283953333241515520 & 156692388 & SDSS\,J142559.17+284715.2 & 16.7380 & 23 & 1.4896 & 2 objects within 30" \\
& & & & & & \gaia\ DR2\,382086995098288256 \\
\rowcolor{white}
\rowcolors{25}{}{light3}
\multirow{-2}{*}{382086995098288896}& \multirow{-2}{*}{440171028} & \multirow{-2}{*}{FBS\,0021+418} & \multirow{-2}{*}{15.7920} & \multirow{-2}{*}{17} & \multirow{-2}{*}{0.1935} & \citet{drake14} 12" away \\
\rowcolor{light3}
880252005422941440 & 4161582 & LAMOST\,J073756.25+311646.5 & 13.5824 & 20 & 0.2574 & RR\,Lyr, \citet{drake14} \\
1302698627809988096 & 459285617 & PG\,1610+239A & 13.0425 & 24,25 & 1.9579 & - \\
130950357400044800 & 620408944 & KUV\,02226+2835 & 17.3505 & 18 & 2.3714 & 4 bright objects within 53" \\
1382933122321062912 & 29385876 & FBS\,1554+403 & 14.0505 & 24 & 0.9382 & - \\
1502297272862648704 & 393911299 & BSD\,33-110 & 11.3418 & 16,22,23 & 1.2393 & - \\
2135353702585217920 & 27766711 & KIC\,12021724 & 15.4766 & 14,15 & 0.3879 & - \\
3922570889586104064 & 86277081 & PG\,1206+165 & 13.6527 & 22 & 0.9000 & - \\
915310139832816256 & 27319282 & PG\,0825+428 & 15.0510 & 20 & 1.2956 & - \\
\hline
\rowcolor{light2}
\multicolumn{7}{c}{variable sdBs -- Figure\,\ref{fig:sdBf}} \\
\hline
1075984763296931456 & 313355841 & PG\,1155+741 & 15.3427 & 14,20,21 & 0.8691 & - \\
1270423258548105728 & 357500534 & PG\,1517+265 & 15.8639 & 24 & 1.0576 & - \\
1282064372266357888 & 1101440486 & SDSS\,J145634.65+300450.9 & 16.7505 & 24 & 0.4367 & - \\
1313140659676082560 & 298336617 & PG\,1648+315 & 15.8415 & 25 & 1.0069 & - \\
1380875729907633408 & 1200857883 & KUV\,16160+4120 & 17.0611 & 24,25 & 1.0133 & - \\
1394116942282089344 & 155947880 & PG\,1524+439 & 15.0667 & 23,24 & 1.5255 & - \\
142389199635140864 & 68156854 & FBS\,0255+379 & 14.6431 & 18 & 0.9211 & - \\
1434391159854454144 & 233683336 & PG\,1723+603 & 15.5060 & 14-17,19-26 & 0.25 - 2 & - \\
1629425796564620160 & 198176924 & HS\,1615+6341 & 15.8057 & 14-26 & 1.7533 & - \\
1644174130142213632 & 1102460173 & GALEX\,J15134+6454 & 19.6992 & 14-16,21-23 & 3.3196 & - \\
1702141998067113216 & 257002616 & TYC\,4559-2508-1 & 10.1276 & 14-15,20-22,26 & 2.1527 & - \\
1816969872879974656 & 219647492 & KPD\,2022+2033 & 13.6885 & 14 & 1.13872 & crowded field \\
1904789240973843072 & 128784655 & FBS\,2238+369 & 14.1718 & 16 & 1.9962 & - \\
1949954704743455104 & 256785606 & FBS\,2155+374 & 14.1748 & 15,16 & 0.3534 & 3 objects within 21" \\
1966389723521001728 & 372181885 & LAMOST\,J213853.04+402415.7 & 14.4891 & 15,16 & 0.8289 & object 20" away \\
2114190543287369856 & 193944889 & FBS\,1801+431 & 14.3941 & 25 & 0.5566 & - \\
2129739699292289152 & 26492416 & Kepler\,J192652+490849 & 15.3157 & 14,15 & 3.0976 & \citet{ostensen11} \\
2809421366255514112 & 301799840 & PG\,0033+266 & 14.2751 & 17 & 0.2180 & - \\
346721341030064640 & 291885456 & FBS\,0156+439 & 15.3142 & 18 & 1.4519 & 4 objects within 21" \\
3744179869623125888 & 138173268 & Balloon\,81300002 & 13.8032 & 23 & 1.9374 & - \\
375833419635317888 & 238627422 & FBS\,0048+432A & 14.6036 & 17 & 1.4989 & - \\
380338531092183424 & 288292938 & LAMOST\,J002124.79+402857.1 & 15.5119 & 17 & 0.7905 & crowded field \\
385493350841270528 & 440072535 & FBS\,0013+434 & 15.5555 & 17 & 0.2600 & - \\
751830524767171200 & 450325981 & CBS\,129 & 11.2693 & 21 & 0.3 - 10 & - \\
833223213044650240 & 406756832 & PG\,1022+459 & 15.8435 & 21 & 2.4213 & - \\
\hline
\end{tabular}}
\end{table*}

\subsection{Spectroscopically confirmed other subdwarf variables}
\begin{figure*}
\includegraphics[width=1\textwidth]{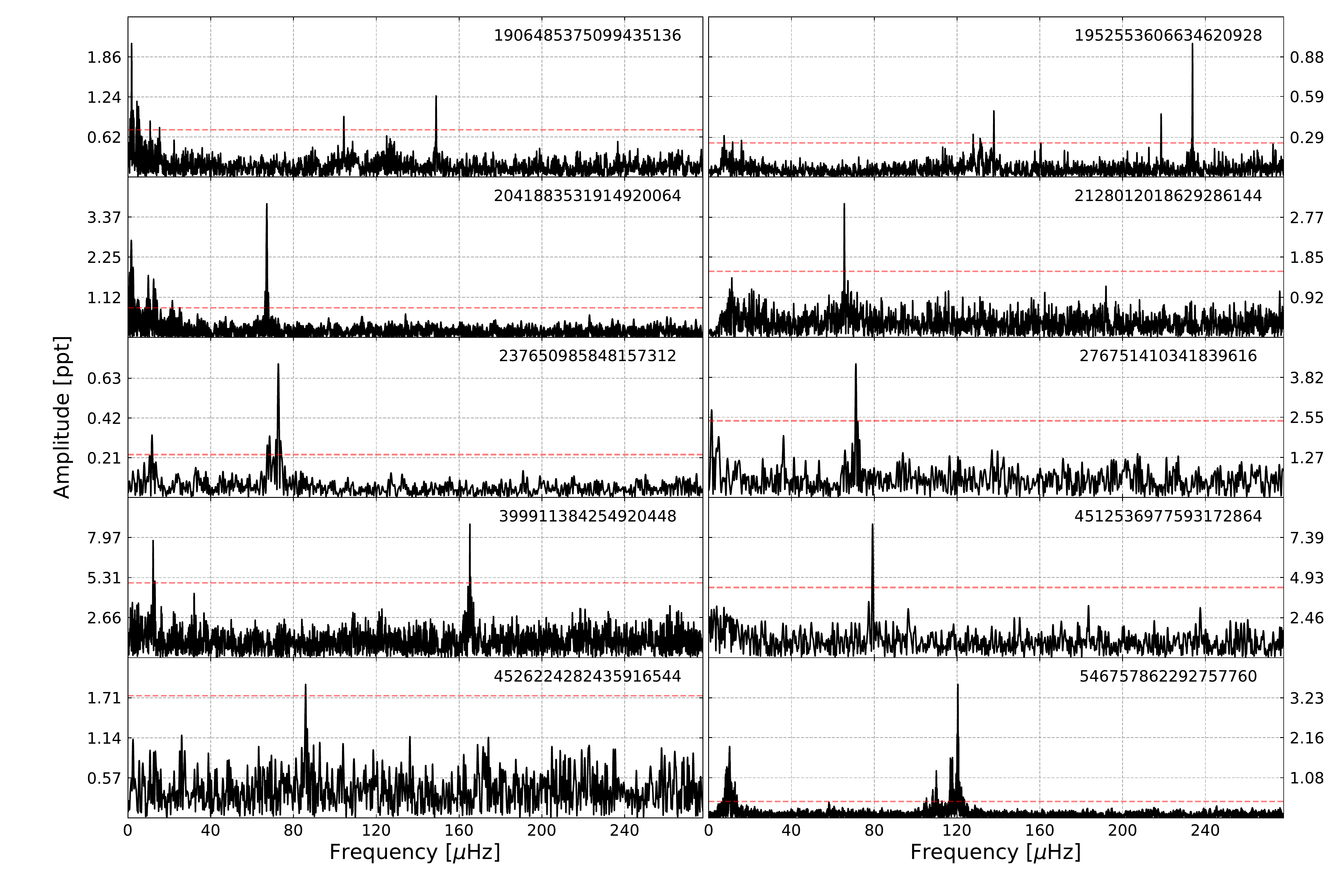}
\caption{Amplitude spectra of 10 sd variable stars listed in the first group of Table\,\ref{tab:sd}.}
\label{fig:sdV}
\end{figure*}

\begin{figure*}
\includegraphics[width=1\textwidth]{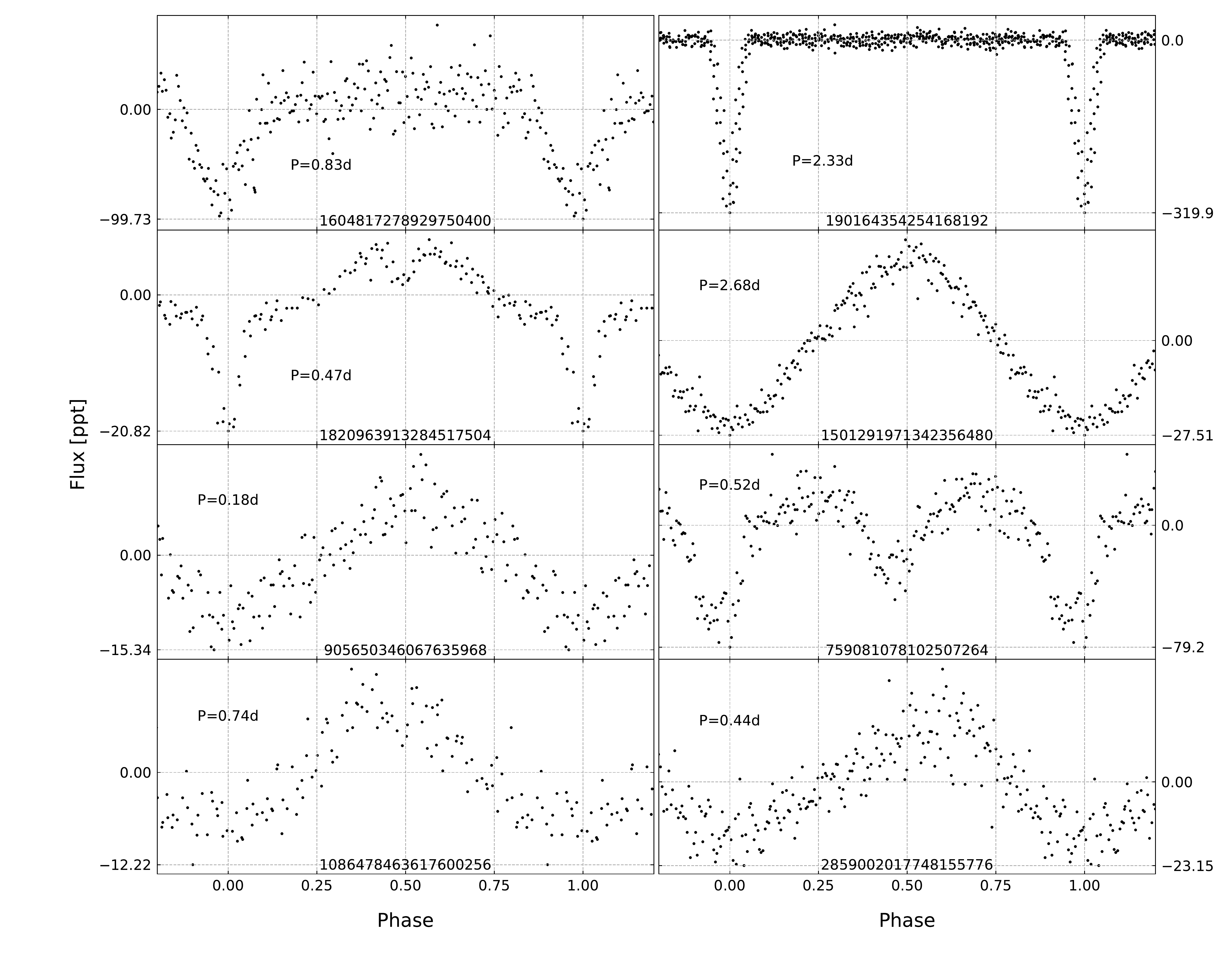}
\caption{Phased and binned time-series data of variable sd stars listed in the second group of Table\,\ref{tab:sd}.}
\label{fig:sdp}
\end{figure*}

\begin{figure*}
\includegraphics[width=1\textwidth]{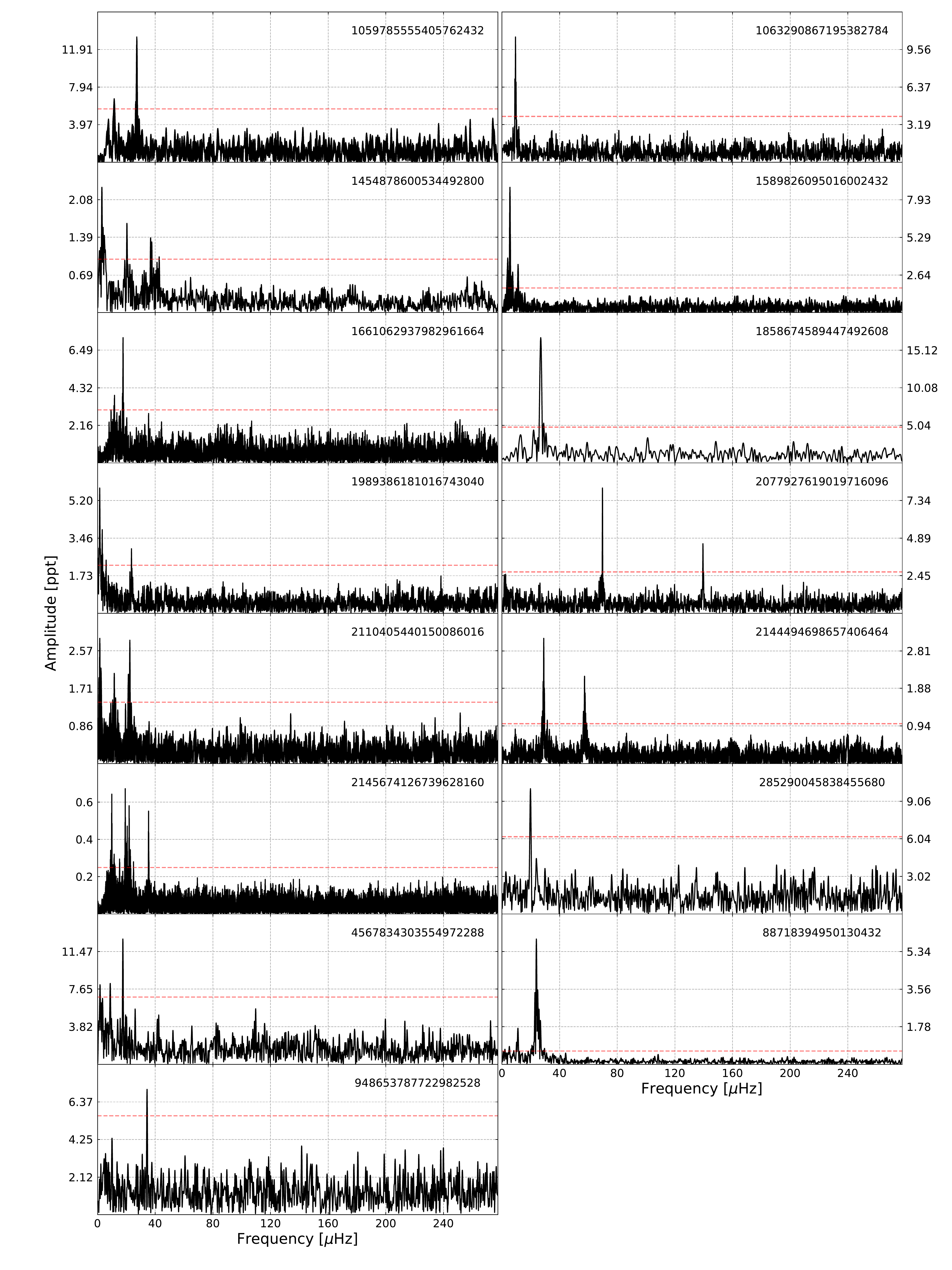}
\caption{Amplitude spectra of variable sd stars listed in the third group of Table\,\ref{tab:sd}.}
\label{fig:sdf}
\end{figure*}

We found 33 objects, which are classified by \citet{geier20} as subdwarfs, showing flux variations similar to the objects reported in Section\,\ref{sec:sdB}. These objects are not classified as sdBs but as sdO (14), sdOB (6), or just sd (13). The objects of the latter class require further classification to confirm the spectral type to O, B or OB. We list these objects in Table\,\ref{tab:sd}. It is separated into three groups in the same way as sdBs objects in Section\,\ref{sec:sdB}. Most of the objects in the first group, {\it i.e.} likely pulsating sds, are single peaks, with \gaia\ DR2\,1952553606634620928 being the richest among all sds and sdBs, having five peaks above our threshold of S/N\,=\,4.5. We show the objects of this group in Figure\,\ref{fig:sdV}. The first three objects in the second group (Figure\,\ref{fig:sdp}) are eclipsing binaries, characterized by sharp eclipses, with \gaia\ DR2\,1820963913284517504 showing an \hwvir\ shape of the phased time-series data, and having one of the longest orbital period among as compared to other \hwvir\ systems. The system was found to be planetary nebula, so the primary component is not an extreme horizontal branch star. Next two objects in Figure\,\ref{fig:sdp} show one symmetric maxima, the following object shows two maxima (eclipsing systems) and the two remaining ones show one asymmetric maxima. \gaia\ DR2\,759081078102507264 has been previously found to be variable by \cite{drake14}. The third group contains objects selected based on the same criteria as in Section\,\ref{sec:sdB}. The majority of these are most likely binaries and we plot their amplitude spectra in Figure\,\ref{fig:sdf}.

\begin{table*}
\centering
\caption{Basic information of 33 objects classified as sds. Time-series data or amplitude spectra of these objects are plotted in Figs.\,\ref{fig:sdV},\ref{fig:sdp} and \ref{fig:sdf}.}
\label{tab:sd}
\rowcolors{1}{light3}{}
\resizebox{0.85\textwidth}{!}{\begin{tabular}{lrlcccl}
\hline
\rowcolor{light1}
& & & \multicolumn{1}{c}{G} & & \multicolumn{1}{c}{Period} & \\
\rowcolor{light1}
\multicolumn{1}{c}{\multirow{-2}{*}{\gaia\ DR2}} & \multicolumn{1}{c}{\multirow{-2}{*}{TIC}} & \multicolumn{1}{c}{\multirow{-2}{*}{Name}} & \multicolumn{1}{c}{[mag]} & \multicolumn{1}{c}{\multirow{-2}{*}{Sector}} & \multicolumn{1}{c}{[days]} &  \multicolumn{1}{c}{\multirow{-2}{*}{Remarks}}\\
\hline
\rowcolor{light2}
\multicolumn{7}{c}{sdVs -- Figure\,\ref{fig:sdV}}\\
\hline
1906485375099435136 & 259091223 & FBS2209+354 & 14.3005 & 15,16 & 0.07-0.13 & - \\
1952553606634620928 & 407657360 & LAMOSTJ214600.31+372119.7 & 14.6625 & 15,16 & 0.045 - 0.1 & 4 objects within 32" \\
2041883531914920064 & 20688004 & GALEXJ18578+3048 & 13.7315 & 14,26 & 0.1726 & 2 objects within 21" \\
2128012018629286144 & 1882679963 & KeplerJ19352+4555 & 17.1559 & 14,15 & 0.1766 & - \\
237650985848157312 & 194781979 & LAMOSTJ032717.71+410344.5 & 10.1918 & 18 & 0.1594 & - \\
276751410341839616 & 372463918 & SDSSJ041536.69+560222.5 & 14.3548 & 19 & 0.1627 & - \\
399911384254920448 & 354155298 & SDSSJ012458.96+475640.9 & 16.9215 & 17,18 & 0.0701 & 2 objects within 32" \\
4512536977593172864 & 346754900 & SDSSJ184336.23+183541.2 & 15.5500 & 26 & 0.1462 & crowded field \\
4526224282435916544 & 1684677537 & LAMOSTJ181102.81+173759.8 & 17.1021 & 26 & 0.1349 & 4 objects within 21" \\
\rowcolors{13}{}{light3}
546757862292757760 & 396013351 & SDSSJ022718.01+733611.1 & 14.8957 & 18,19,25 & 0.09 - 0.2 & 3 objects within 21" \\
\hline
\rowcolor{light2}
\multicolumn{7}{c}{variable sds -- Figure\,\ref{fig:sdp}}\\
\hline
1604817278929750400 & 1001259992 & SDSSJ142017.21+513904.1 & 18.9939 & 16,22,23 & 0.8253 & - \\
190164354254168192 & 426264139 & LAMOSTJ054257.76+391151.2 & 14.9411 & 19 & 2.3316 & 3 objects within 21" \\
& & & & & & \citet{bond78}, \\
\rowcolor{white}
\multirow{-2}{*}{1820963913284517504} & \rowcolors{25}{}{white}
\multirow{-2}{*}{1842385646} & \multirow{-2}{*}{PNA\,6663} & \multirow{-2}{*}{15.0966} & \multirow{-2}{*}{14} & \multirow{-2}{*}{0.4653} & very crowded field \\
1501291971342356480 & 1001024276 & SDSSJ133627.35+422910.90 & 18.5819 & 22,23 & 2.6765 & object 24" away \\
905650346067635968 & 139150535 & SDSSJ080327.93+342140.7 & 15.0617 & 20 & 0.1824 & object 19" away \\
759081078102507264 & 144353046 & FBS1125+345 & 16.1367 & 22 & 0.5228 & \citet{drake14} \\
1086478463617600256 & 85210423 & GALEXJ07354+6132 & 13.9744 & 20 & 0.7398 & - \\
2859002017748155776 & 58105203 & PG0023+298 & 15.0708 & 17 & 0.4415 & - \\
\hline
\rowcolor{light2}
\multicolumn{7}{c}{variable sds -- Figure\,\ref{fig:sdf}}\\
\hline
\rowcolor{white}
\rowcolors{26}{light3}{}
1059785555405762432 & 103802941 & PG1030+665 & 15.4705 & 14,21 & 0.4251 & - \\
1063290867195382784 & 86374556 & SDSSJ093654.17+621449.5 & 15.8320 & 20,21 & 1.2309 & - \\
1454878600534492800 & 199497852 & vZ1128 & 14.9662 & 23 & 0.5684 & crowded field \\
1589826095016002432 & 116307277 & PG1447+459 & 14.9342 & 16,23 & 2.0965 & 2 objects within 21" \\
1661062937982961664 & 1001364570 & PG1348+606 & 16.3181 & 15,16,22 & 0.6538 & - \\
1858674589447492608 & 230775376 & Feige114 & 13.5045 & 15 & 0.4302 & 4 bright objects within 21" \\
1989386181016743040 & 252589558 & KPD2259+5149 & 13.8445 & 16,17 & 0.4920 & crowded field \\
2077927619019716096 & 271345879 & KPD1938+4220 & 15.5654 & 14,15 & 0.1660 & 4 objects within 21" \\
2110405440150086016 & 157583405 & SDSSJ183620.82+405938.5 & 14.5125 & 14,26 & 0.5 - 10 & object 13" away \\
2144494698657406464 & 48191737 & O11J185046+510738 & 13.6951 & 14,15,26 & 0.3990 & - \\
2145674126739628160 & 47670365 & O11J183634+531657 & 12.4810 & 14,15,19,22,25,26 & 0.31 - 2 & 2 objects within 21" \\
285290045838455680 & 668928634 & SDSSJ045912.47+605156.6 & 18.7139 & 19 & 0.5851 & - \\
4567834303554972288 & 471013471 & PG1707+214 & 15.6066 & 25 & 1.3179 & 3 bright objects within 3 \\
88718394950130432 & 426419494 & SDSSJ024113.47+215743.2 & 12.7945 & 18 & 0.9694 & - \\
948653787722982528 & 742179265 & SDSSJ072401.73+410320.9 & 19.3820 & 20 & 0.3371 & 4 objects within 32" \\
\hline
\end{tabular}}
\end{table*}

\subsection{Spectroscopically unconfirmed sdBV candidates}
\label{pulsators}
\begin{figure*}
\includegraphics[width=0.85\textwidth]{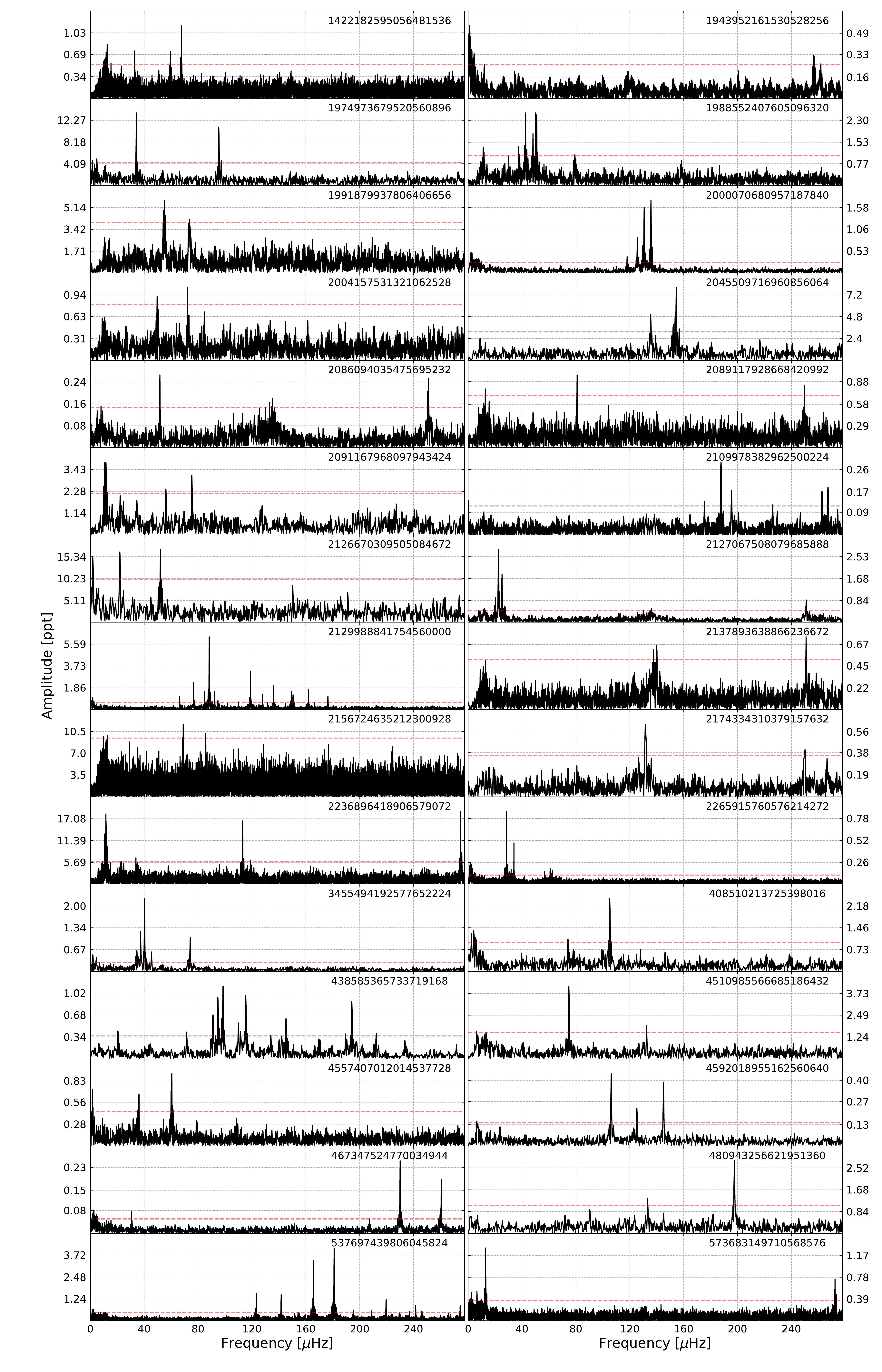}
\caption{Amplitude spectra of sdBV candidates that are not spectroscopically classified listed in Table\,\ref{tab:pulsators}.}
\label{fig:pulsators_ft}
\end{figure*}

We found 30 objects that show multi-unrelated-peak amplitude spectra, which we interpret as pulsations. We show the amplitude spectra of these objects in Figure\,\ref{fig:pulsators_ft}. The spectra show peaks in the g-mode region of sdBVs and that is why these objects are of particular interest to us. These stars may also be $\delta$~Scuti stars, though \citet{geier20} applied a color index criterion to avoid cool stars, or $\beta$\,Cep stars. \citet{geier20} provides detailed arguments using \gaia\ color indices that these targets are hot subluminous stars and occupy the region -0.7 < G$_{\rm BP}$-G$_{\rm RP} \lesssim$ 0.7 in the \gaia\ colour space. These objects are not spectroscopically classified so we are unable to make any definite conclusion about their pulsation nature, however the amplitude spectra are richer in peaks that any of likely sdBVs or sdVs we show in Figures\,\ref{fig:sdBV} and \ref{fig:sdV}. We listed these objects along with their basic information in Table\,\ref{tab:pulsators}. Of special interest are \gaia\ DR2\,2129988841754560000, \gaia\ DR2\,438585365733719168 and \gaia\ DR2\,537697439806045824 since their spectra are the richest, containing enough peaks to make a mode identification by means of period spacings. Even though we are not positive the three objects are sdBVs, our suggested mode assignment may be useful if a sdB classification is confirmed by future spectroscopic analyses. We present the result of the mode identification in Section\,\ref{sec:modeID}. \gaia\ DR2\,2126670309505084672 was found to be a variable by \citet{reinhold13} and denoted KIC\,9020774. The optimal aperture for \gaia\ DR2\,2127067508079685888 overlaps with \gaia\ DR2\,2127067542439425536 (KIC\,8879964) whose variability with the same period has been reported by \citet{reinhold13}. It is most likely that the variability comes from the latter object.

\begin{table*}
\centering
\caption{Basic information of 30 sdBV candidates that are not spectroscopically classified. We show amplitude spectra of these objects in Figure\,\ref{fig:pulsators_ft}}
\label{tab:pulsators}
\rowcolors{1}{}{light3}
\resizebox{0.85\textwidth}{!}{\begin{tabular}{lrccl}
\hline
\rowcolor{light1}
& & \multicolumn{1}{c}{G} & & \\
\rowcolor{light1}
\multicolumn{1}{c}{\multirow{-2}{*}{\gaia\ DR2}} & \multicolumn{1}{c}{\multirow{-2}{*}{TIC}} & \multicolumn{1}{c}{[mag]} & \multicolumn{1}{c}{\multirow{-2}{*}{Sector}} & \multicolumn{1}{c}{\multirow{-2}{*}{Remarks}}\\
\hline
1422182595056481536 & 320525680 & 13.8708 & 14-21,23-26 & 2 objects within 21" \\
1943952161530528256 & 431548978 & 15.6118 & 17,24 & - \\
1974973679520560896 & 311792028 & 15.8844 & 16 & 4 objects within 21" \\
1988552407605096320 & 66784300 & 14.8772 & 16,17 & 5 bright objects within 21" \\
1991879937806406656 & 2044241813 & 18.4712 & 16,17 & 3 objects within 21" \\
2000070680957187840 & 2013748140 & 17.9850 & 16,17 & crowded field \\
2004157531321062528 & 2014779767 & 17.4638 & 16,17 & crowded field \\
2045509716960856064 & 1873643239 & 17.8928 & 14 & 2 bright objects within 21" \\
2086094035475695232 & 1881839953 & 17.9128 & 14,15 & crowded field \\
2089117928668420992 & 364910983 & 16.2671 & 14,15,16 & 2 bright objects within 21" \\
2091167968097943424 & 1550027150 & 17.4874 & 26 & 3 objects within 21" \\
2109978382962500224 & 1550453189 & 17.2795 & 14,26 & - \\
2126670309505084672 & 159722705 & 15.0028 & 15 & \citet{reinhold13} \\
2127067508079685888 & 159108456 & 16.3927 & 14,15 & KIC\,8879964, \citet{reinhold13} 9" away \\
2129988841754560000 & 1882977987 & 16.5499 & 14,15 & crowded field \\
2137893638866236672 & 1883445550 & 18.2584 & 14,15,16 & - \\
2156724635212300928 & 233607898 & 16.0483 & 14-17,19-26 & object 10" away \\
2174334310379157632 & 2017774400 & 16.2330 & 16,17 & crowded field \\
2236896418906579072 & 236868718 & 17.3690 & 14-17,24 & - \\
2265915760576214272 & 229786221 & 13.4680 & 14-17,19-23,25-26 & - \\
3455494192577652224 & 116747928 & 11.4189 & 19 & - \\
408510213725398016 & 623262377 & 18.6718 & 18 & 3 objects within 21" \\
438585365733719168 & 428055301 & 15.5791 & 18 & crowded field \\
4510985566685186432 & 298514603 & 15.4720 & 26 & crowded field \\
4557407012014537728 & 311431337 & 12.0714 & 25,26 & - \\
4592018955162560640 & 24261156 & 16.0873 & 26 & crowded field \\
467347524770034944 & 51026936 & 16.4141 & 18,19 & 5 bright objects within 32" \\
480943256621951360 & 328663453 & 16.0841 & 19 & object 13" away \\
537697439806045824 & 407425099 & 16.4121 & 18,19,24,25 & crowded field \\
573683149710568576 & 461640603 & 14.2553 & 18,20,25,26 & 2 bright objects within 21" \\
\hline
\end{tabular}}
\end{table*}

\subsection{Spectroscopically unclassified variables}
This sample includes all objects that are not spectroscopically classified and do not show amplitude spectra typical for sdBs.

\begin{figure*}
\includegraphics[width=1\textwidth]{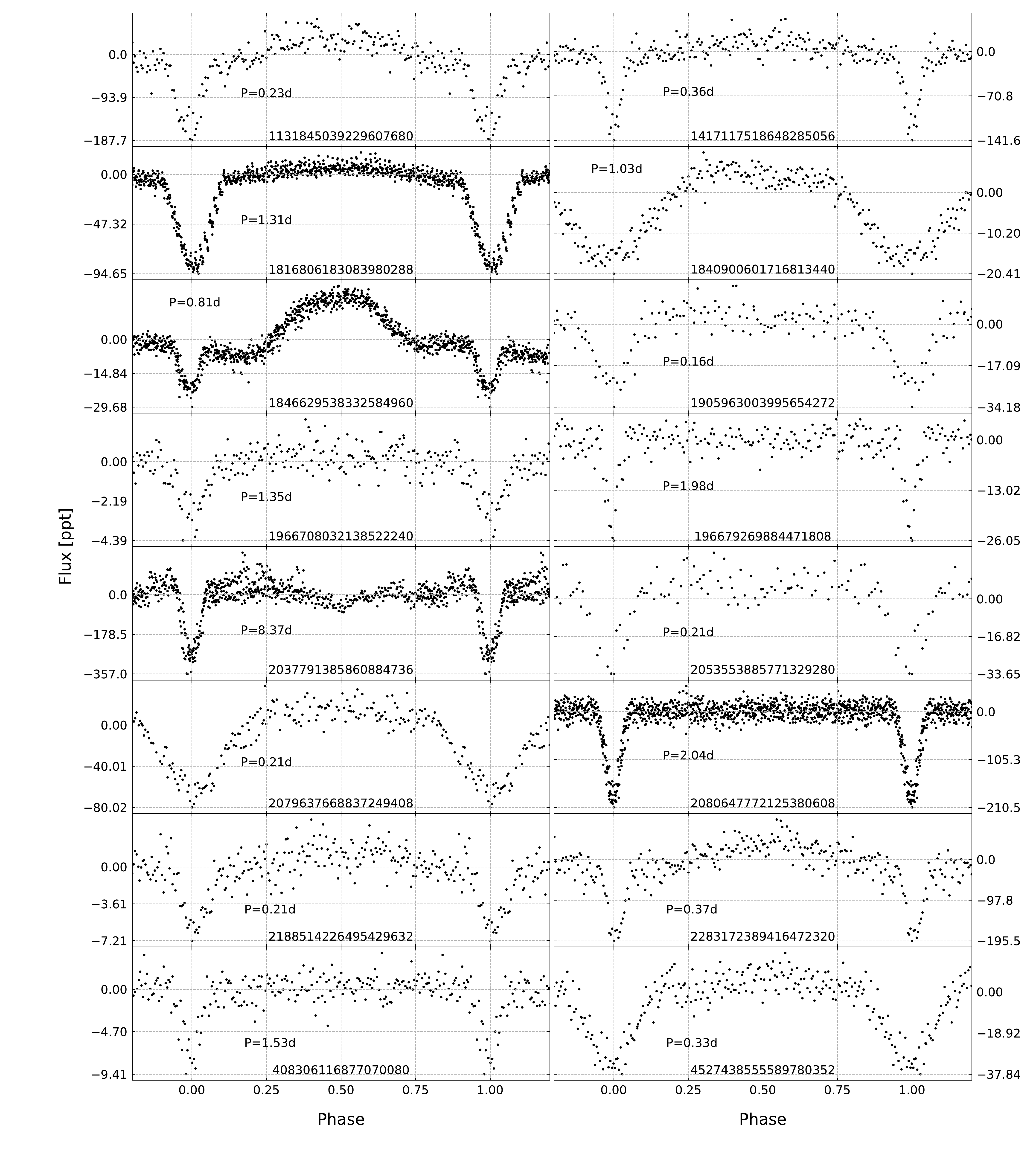}
\caption{Phased time-series data of eclipsing binaries that show sharp primary eclipses only, and are listed in the first group of Table\,\ref{tab:eb}.}
\label{fig:eb_pr}
\end{figure*}

\begin{figure*}
\includegraphics[width=1\textwidth]{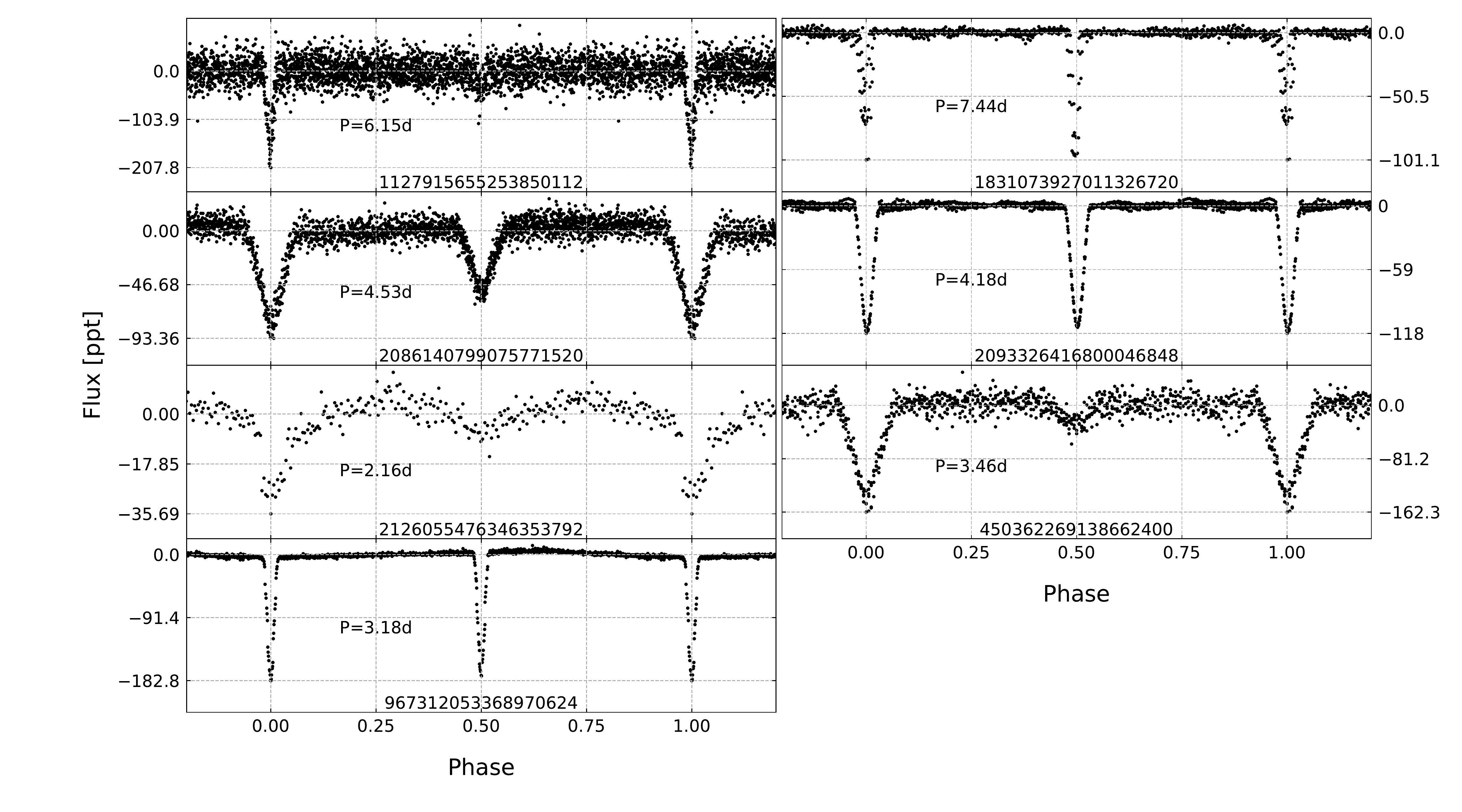}
\caption{Phased time-series data of eclipsing binaries that show sharp both primary and secondary eclipses, and are listed in the second group of Table\,\ref{tab:eb}.}
\label{fig:eb_ps}
\end{figure*}

We found 23 eclipsing binaries with sharp eclipses that are not yet spectroscopically classified. They all show distinct eclipses and are either detached or semi-detached binaries. We report the entire list of these objects in Table\,\ref{tab:eb}. Possible contact binaries are not included in this group. We separated these 23 objects into two groups. The first group consists of 16 objects that show only primary eclipses. This may be a consequence of a low inclination angle and/or small size of primary components with respect to the distance between them. We show phased time-series data of these objects in Figure\,\ref{fig:eb_pr}. There are five candidates for \hwvir\ systems. They show no detectable secondary eclipses but they show a flux increase between primary eclipses that is characteristic of a reflection effect. None of these five objects has been reported thus far. The second group consists of seven objects that show both primary and secondary eclipses. We show phased time-series data of these objects in Figure\,\ref{fig:eb_ps}. \gaia\ DR2\,2126055476346353792 has been previously identified as a cataclysmic variable by \citet{scaringi13} and denoted KIC\,7524178.

\begin{table*}
\centering
\caption{Basic information of 23 spectroscopically unclassified eclipsing binaries. We show the time-series data of these objects in Figures\,\ref{fig:eb_pr} and \ref{fig:eb_ps}.}
\label{tab:eb}
\rowcolors{1}{light3}{}
\resizebox{0.85\textwidth}{!}{\begin{tabular}{lrcccl}
\hline
\rowcolor{light1}
& & \multicolumn{1}{c}{G} & & \multicolumn{1}{c}{Period} & \\
\rowcolor{light1}
\multicolumn{1}{c}{\multirow{-2}{*}{\gaia\ DR2}} & \multicolumn{1}{c}{\multirow{-2}{*}{TIC}} & \multicolumn{1}{c}{[mag]} & \multicolumn{1}{c}{\multirow{-2}{*}{Sector}} & \multicolumn{1}{c}{[days]} & \multicolumn{1}{c}{\multirow{-2}{*}{Remarks}}\\
\hline
\rowcolor{light2}
\multicolumn{6}{c}{Only primary eclipses detected -- Figure\,\ref{fig:eb_pr}}\\
\hline
1131845039229607680 & 459182998 & 16.1614 & 14,26 & 0.2344 & \hwvir\ candidate \\
1417117518648285056 & 1400704733 & 17.0325 & 14-15,17-18,20-21,23-26 & 0.3637 & \hwvir\ candidate \\
1816806183083980288 & 1943324398 & 17.2206 & 14 & 1.3135 & crowded field \\
1840900601716813440 & 1951174238 & 18.9249 & 15 & 1.0329 & 3 objects within 21" \\
1846629538332584960 & 15040115 & 11.8285 & 15 & 0.8099 & 3 objects within 32" \\
1905963003995654272 & 274852575 & 16.9924 & 16 & 0.1610 & bright object 7" away \\
1966708032138522240 & 372019916 & 15.4010 & 15,16 & 1.3454 & 2 objects within 21" \\
196679269884471808 & 701334595 & 17.7576 & 19 & 1.9838 & 2 objects within 21" \\
2037791385860884736 & 1712396254 & 18.6493 & 14 & 8.3660 & crowded field \\
2053553885771329280 & 137755255 & 15.4784 & 14 & 0.2060 & 3 objects within 21" \\
2079637668837249408 & 1881471898 & 17.4596 & 14,15 & 0.2075 & 4 objects within 21" \\
2080647772125380608 & 1881706041 & 18.5749 & 14,15 & 2.0397 & 3 bright objects within 32" \\
& & & & & \hwvir\ candidate \\
\rowcolor{white}
\rowcolors{17}{}{light3}
\multirow{-2}{*}{2188514226495429632} & \multirow{-2}{*}{1979105817} & \multirow{-2}{*}{17.5747} & \multirow{-2}{*}{16,17} & \multirow{-2}{*}{0.2053} & 5 objects within 21" \\
2283172389416472320 & 2051607908 & 17.6481 & 18,19,24,26 & 0.3682 & \hwvir\ candidate \\
408306116877070080 & 623229903 & 18.9346 & 18 & 1.5349 & crowded field \\
& & & & & \hwvir\ candidate \\
\rowcolor{light3}
\multirow{-2}{*}{4527438555589780352} & \multirow{-2}{*}{1684897611} & \multirow{-2}{*}{17.5813} & \multirow{-2}{*}{26} & \multirow{-2}{*}{0.3295} & 3 objects within 21" \\
\hline
\rowcolor{light2}
\multicolumn{6}{c}{Both primary and secondary eclipses detected -- Figure\,\ref{fig:eb_ps}}\\
\hline
1127915655253850112 & 841356486 & 18.5056 & 14,20,21,26 & 6.1499 & object 12" away \\
1831073927011326720 & 1947381728 & 17.6647 & 14 & 7.4423 & crowded field \\
2086140799075771520 & 1881852480 & 18.5450 & 14,15 & 4.5270 & 3 bright objects within 21" \\
2093326416800046848 & 1715372736 & 10.6998 & 14,26 & 4.1785 & BD+36\,3302 1" away \\
2126055476346353792 & 159448831 & 17.2137 & 14 & 2.1607 & CV, \citet{scaringi13} \\
450362269138662400 & 623842756 & 16.4752 & 18 & 3.4583 & 4 objects within 21" \\
967312053368970624 & 704256138 & 16.7824 & 20 & 3.1849 & - \\
\hline
\end{tabular}}
\end{table*}

\begin{figure*}
\includegraphics[width=1\textwidth]{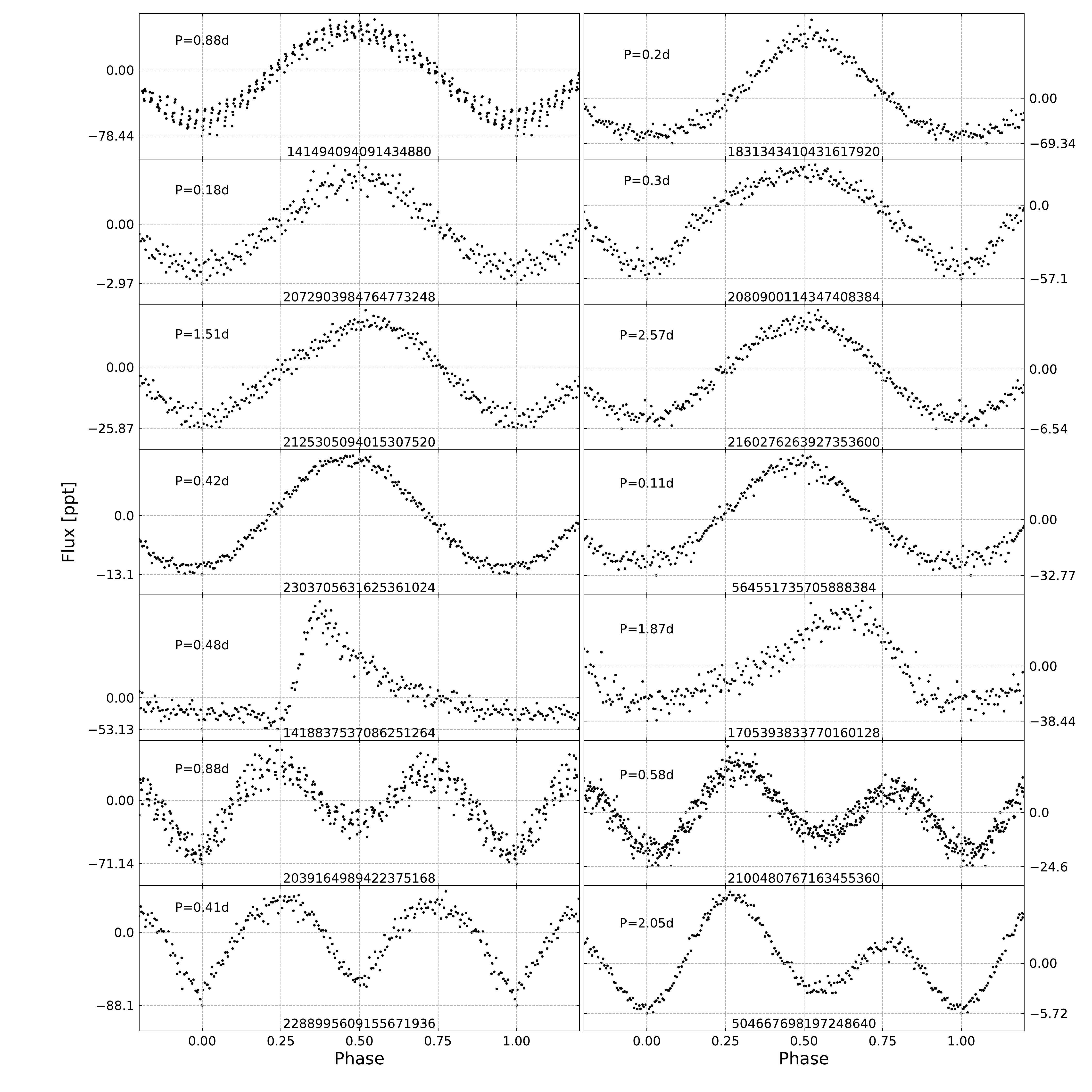}
\caption{Phased time-series data of 14 (out of 93) spectroscopically unclassified binaries without sharp eclipses listed in Table\,\ref{tab:nospecbin}. All plots are included in on-line material only.}
\label{fig:nospecbin}
\end{figure*}

In Table\,\ref{tab:nospecbin} we list a sample of 93 binaries that do not show sharp eclipses and are not spectroscopically classified. We divided the sample into three groups. The first group contains 63 objects that show one symmetric maximum in their phased time-series data. Such maxima can be interpreted {\it e.g.} as a reflection effect observed in binary systems with large temperature difference between two components \citep[for example,][]{baran19}. \gaia\ DR2\,893386457796229504 has been found to be a variable by \citet{drake14}. The second group consists of 16 objects that show one asymmetric maximum in their phased-time series data. Such a shape is characteristic for classical pulsators, which can mean that the selection made by \citet{geier20} is not ideal or color indices are not correct. The third group consists of 14 objects that show two maxima in their phased time-series data. Such a shape may be an indication of contact binaries, which have continuous eclipses, or ellipsoidal binaries. We refer to more details included in Paper I. \gaia\ DR2\,1962740066464303744 is 1" away from V1942\,Cyg, which is a variable star and most likely the source of the variability we find. \gaia\ DR2\,2100480767163455360 was identified as a cataclysmic variable by \citet{fontaine11} and denoted KIC\,4547333. We show a selection of each group in Figure\,\ref{fig:nospecbin}, while plots of all objects are included in on-line material.

\begin{table*}
\centering
\caption{Basic information of 14 (out of 93) spectroscopically unclassified binaries without sharp eclipses. We show the time-series data of these objects in Figure\,\ref{fig:nospecbin}. The full list of the objects in this group is included in on-line material.}
\label{tab:nospecbin}
\rowcolors{1}{light3}{}
\resizebox{0.85\textwidth}{!}{\begin{tabular}{lrcccl}
\hline
\rowcolor{light1}
& & \multicolumn{1}{c}{G} & & \multicolumn{1}{c}{Period} & \\
\rowcolor{light1}
\multicolumn{1}{c}{\multirow{-2}{*}{\gaia\ DR2}} & \multicolumn{1}{c}{\multirow{-2}{*}{TIC}} & \multicolumn{1}{c}{[mag]} & \multicolumn{1}{c}{\multirow{-2}{*}{Sector}} & \multicolumn{1}{c}{[days]} & \multicolumn{1}{c}{\multirow{-2}{*}{Remarks}}\\
\hline
\rowcolor{light2}
\multicolumn{6}{c}{One symmetric maximum detected -- Figure\,\ref{fig:nospecbin}}\\
\hline
141494094091434880 & 67658009 & 13.0436 & 18 & 0.8764 & - \\
1831343410431617920 & 406417817 & 14.7108 & 14 & 0.2010 & crowded field \\
2072903984764773248 & 1879342714 & 17.8580 & 14 & 0.1761 & crowded field \\
2080900114347408384 & 416641307 & 14.5962 & 14,15 & 0.3011 & 3 objects within 21" \\
2125305094015307520 & 21018674 & 13.4058 & 14,15,18,21,25,26 & 1.5079 & - \\
2160276263927353600 & 233733792 & 12.4555 & 14-17,19-26 & 2.5694 & - \\
2303705631625361024 & 397532904 & 12.8886 & 18,19,20,25,26 & 0.4204 & - \\
564551735705888384 & 609725827 & 16.5793 & 18,19,24,25,26 & 0.1068 & 5 objects within 32" \\
\hline
\rowcolor{light2}
\multicolumn{6}{c}{One asymmetric maximum detected -- Figure\,\ref{fig:nospecbin}}\\
\hline
\rowcolor{white}
\rowcolors{13}{}{light3}
1418837537086251264 & 1271027428 & 18.9619 & 17,20,23-26 & 0.4771 & - \\
1705393833770160128 & 288381009 & 16.4297 & 14-26 & 1.8706 & object 13" away \\
\hline
\rowcolor{light2}
\multicolumn{6}{c}{Two maxima detected -- Figure\,\ref{fig:nospecbin}}\\
\hline
\rowcolor{white}
\rowcolors{15}{}{light3}
2039164989422375168 & 1712800054 & 16.6814 & 14 & 0.8764 & crowded field \\
2100480767163455360 & 121107327 & 16.2609 & 14 & 0.5772 & CV, \citet{fontaine11} \\
2288995609155671936 & 1884532373 & 16.5623 & 14-26 & 0.4107 & - \\
504667698197248640 & 445526365 & 10.1176 & 18 & 2.0482 & crowded field \\
\hline
\end{tabular}}
\end{table*}

\begin{figure*}
\includegraphics[width=1\textwidth]{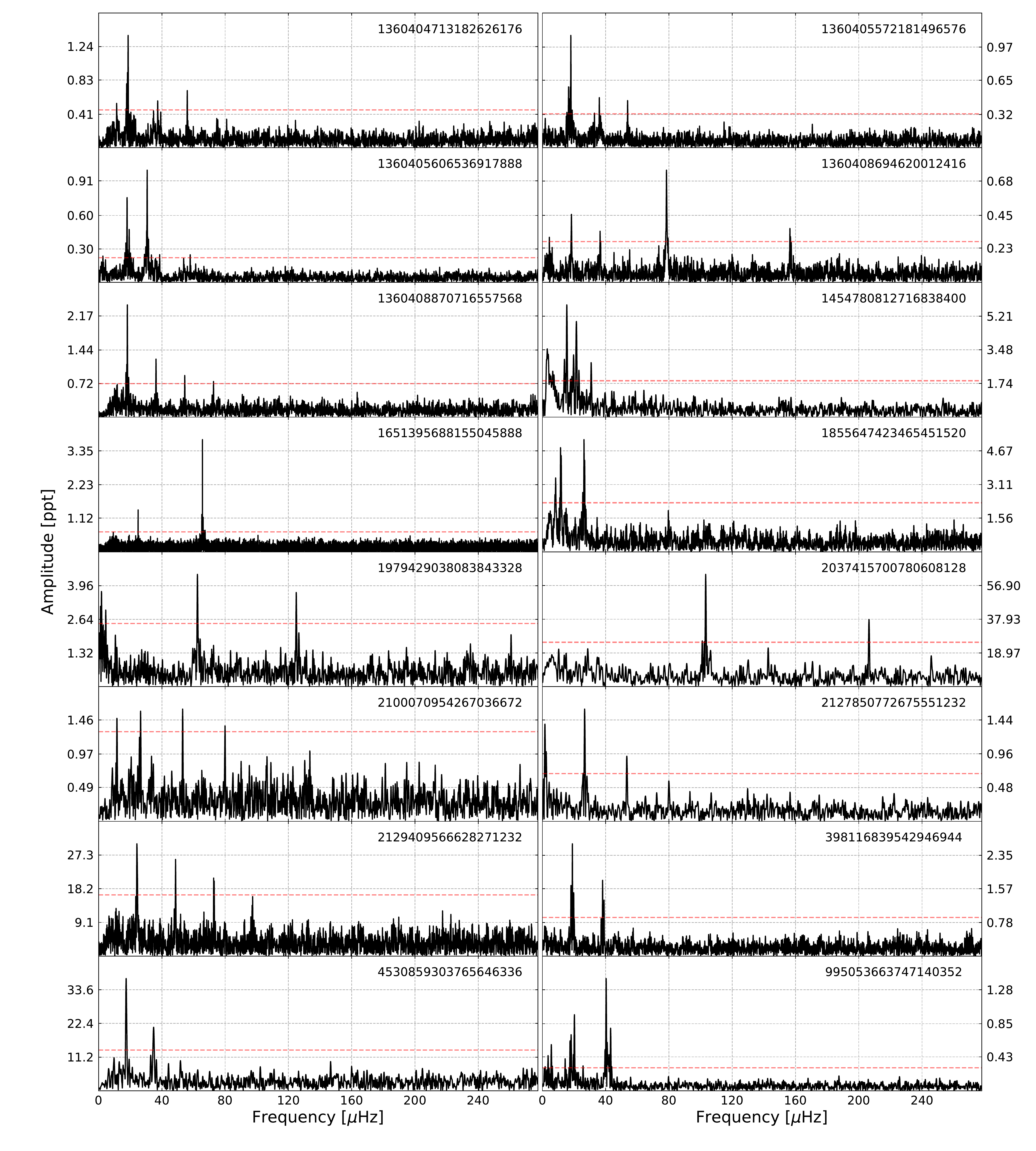}
\caption{Amplitude spectra of 16 (out of 228) spectroscopically unclassified variables listed in Table\,\ref{tab:nospecft}. All plots are included in on-line material only.}
\label{fig:nospecft}
\end{figure*}

We found 228 objects that are not spectroscopically classified and we identified neither as pulsators nor as eclipsing binary systems, presented earlier in this Section. This group includes objects that show peaks consistent with binarity but a small amplitude of a flux variation. The typical S/N is 8 or lower. We find these variations in amplitude spectra. This group also contains targets with multiple unrelated peaks in amplitude spectra, regardless of the S/N, which makes data phasing pointless. This multiplicity of peaks is typical for pulsators, however the amplitude spectra do not resemble the ones of sdBVs, since the unrelated peaks are below 60\,$\upmu$Hz, and that is why we decided not to include them in Section\,\ref{pulsators}. \gaia\ DR2\,2105585421693855744 is identified with the cataclysmic variable V363\,Lyr and reported by \citet{scaringi13}. \gaia\ DR2\,2127444125171341312 is very close to KIC\,9278505 also reported by \citet{scaringi13}. We provided basic information about these objects in Table\,\ref{tab:nospecft}, while we show the amplitude spectra in Figure\,\ref{fig:nospecft}.

\begin{table*}
\centering
\caption{Basic information of 16 (out of 228) spectroscopically unclassified variables that we show in Figure\,\ref{fig:nospecft}. The full list is included in on-line material only.}
\label{tab:nospecft}
\rowcolors{1}{}{light3}
\resizebox{0.85\textwidth}{!}{\begin{tabular}{lrccl}
\hline
\rowcolor{light1}
& & \multicolumn{1}{c}{G} & & \\
\rowcolor{light1}
\multicolumn{1}{c}{\multirow{-2}{*}{\gaia\ DR2}} & \multicolumn{1}{c}{\multirow{-2}{*}{TIC}} & \multicolumn{1}{c}{[mag]} & \multicolumn{1}{c}{\multirow{-2}{*}{Sector}} & \multicolumn{1}{c}{\multirow{-2}{*}{Remarks}}\\
\hline
1360404713182626176 & 1270831583 & 15.8418 & 25,26 & very crowded field \\
1360405572181496576 & 334899867 & 16.2057 & 25,26 & very crowded field \\
1360405606536917888 & 1270837848 & 16.3565 & 25,26 & very crowded field \\
1360408694620012416 & 334899806 & 16.1530 & 25,26 & very crowded field \\
1360408870716557568 & 334899706 & 16.1600 & 25,26 & very crowded field \\
1454780812716838400 & 1000656229 & 18.0202 & 23 & very crowded field \\
1651395688155045888 & 1401049068 & 16.1897 & 14-25 & - \\
1855647423465451520 & 1953033063 & 18.4158 & 14,15 & 2 bright objects within 21" \\
1979429038083843328 & 2011754031 & 15.2450 & 16 & crowded field \\
2037415700780608128 & 1712233248 & 15.6530 & 14 & 3 bright objects within 21" \\
2100070954267036672 & 1715980555 & 17.7709 & 14 & 2 objects within 21" \\
2127850772675551232 & 1882665005 & 17.0644 & 15 & object 19" away \\
2129409566628271232 & 290035516 & 16.0203 & 14,15 & 2 objects within 21" \\
398116839542946944 & 196946103 & 14.4348 & 17,18 & object 16" away \\
4530859303765646336 & 1813173164 & 18.9596 & 26 & 5 objects within 21" \\
995053663747140352 & 444902249 & 10.8428 & 20 & - \\
\hline
\end{tabular}}
\end{table*}

\subsection{Non-sdB classified variables}
Even though some of the objects in \citet{geier20} are classified as non-sdB, we have included these in our search as well. These were first considered candidates for hot stars, but were identified as non-sdB objects, mostly O, B or A main sequence stars or white dwarfs. In total, we found 46 non-sdB variables. The same categories of flux variations emerged, with pulsator candidates, eclipsing binaries, reflection effect binaries, ellipsoidal variables, and classical pulsators, with the remaining objects having variability in their amplitude spectra. We present it likewise, {\it i.e.} pulsator candidates, eclipsing, reflection, ellipsoidal binaries, classical pulsator candidates, remaining objects with variability reported in their amplitude spectra. A table and figures of these objects are included in the on-line material only. The table includes basic information on each object along with additional references or contaminating objects, if any.

We found seven pulsator candidates with one being known before, {\it i.e.} \gaia\ DR2\,2077737678383889408 reported by \citet{reinhold13}. Among three eclipsing binaries with sharp eclipses, \gaia\ DR2\,4322472232203849856, besides primary eclipse shows an additional flux drop between phase 0.75 and 1.0. The drop changes the phase, so it may be either two different drops or one drop that shifts from orbit to orbit. The phased time-series data do not show a secondary eclipse at phase 0.5, though a bit of a downward scatter exists. If it is a sign of the secondary eclipse the additional drop must be caused by a tertiary component. If, instead, the drop is the secondary eclipse, it shifts very quickly as a result of an apsidal motion. The extra drops phases well with both a period of 7.05325\,days and its double. The Simbad database lists this object as a double or multiple star. A follow up photometry span over multiple orbits, should resolve this ambiguity. We found five more objects reported previously as variables. Four of them denoted KIC numbers have been observed with the the \kep\ spacecraft and reported by \cite{reinhold13}, while the last one is identified as an RS\,CVn variable by \citet{drake14}.

\section{Mode identification}
\label{sec:modeID}
We selected three objects where we are able to identify the pulsation modes by means of a rotational splitting and evenly spaced, in period, radial overtones, assuming these are g-mode sdBVs. The objects are the following, \gaia\ DR2\,2129988841754560000, \gaia\ DR2\,438585365733719168 and \gaia\ DR2\,537697439806045824, and were selected since they have the richest amplitude spectra among all pulsator candidates we found.

\begin{figure*}
\includegraphics[width=0.85\textwidth]{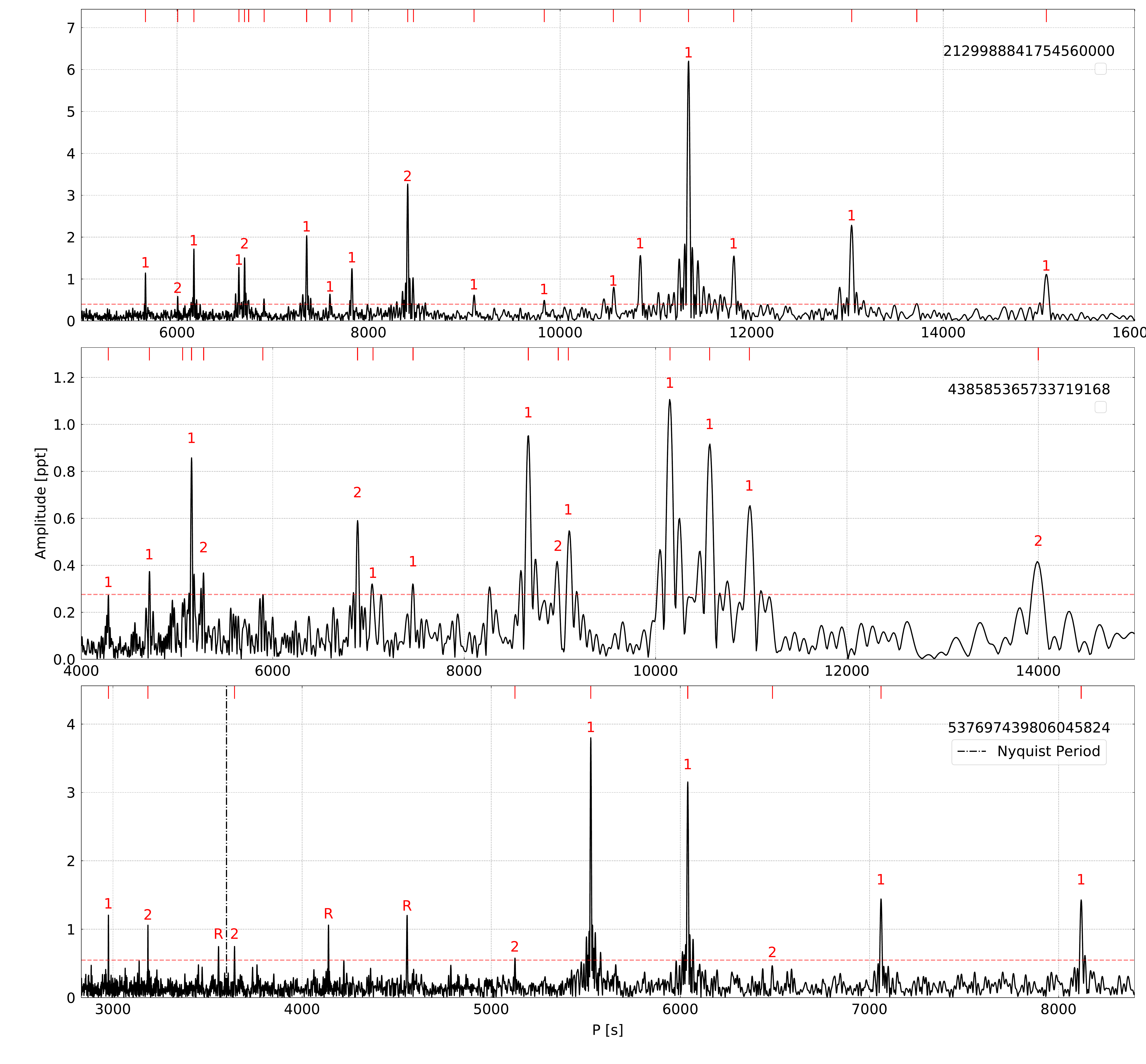}
\caption{Amplitude spectra of three pulsators plotted in period. The red horizontal dashed lines denote $5\,\sigma$ detection threshold, which is consistent with FAP\,=\,0.1\% The modal degrees are shown on top of each detected peak. `R' refers to reflection peaks across the Nyquist period.}
\label{fig:modeids}
\end{figure*}

We followed a standard prewhitening procedure by calculating an amplitude spectrum and removing consecutive peaks by fitting $A_i\sin(2{\pi}f_it+{\phi_i})$ using a non-linear least-square method, where A$_i$ is an amplitude, f$_i$ is a frequency and $\phi_i$ is a phase of an {\it i-th} peak. We used our custom scripts for prewhitening. We removed all peaks down to a detection threshold of about S/N\,=\,5. We updated this threshold level after each peak removal, hence the final threshold was calculated from the residual amplitude spectra, {\it i.e.} with all significant peaks removed.
We present the lists of frequencies detected in each star in Tables\,\ref{tab:modeid1}, \ref{tab:modeid2}, and \ref{tab:modeid3}, while we show the amplitude spectra of these three stars in Figure\,\ref{fig:modeids}.

Since stellar rotation splits pulsation modes into 2{\it l}+1 components of different {\it m} values, and often called multiplets, this splitting can be used for assigning the degree {\it l} of the split modes. One of the first examples with convincing rotationally split modes are \citet{baran09,baran12a,baran12c}. Besides the modal degree identification, a splitting itself is a direct measure of stellar rotation. These are two main reasons why the multiplets are so crucial for mode identification. Multiplets in sdBVs are not easily detectable. A common rotation period of sdBs is around 40\,days, which means that data should cover at least 40\,days, though preferentially 60\,days, as was shown by \citet{baran12c}. The three objects in our sample are two sector data only so our null result search for multiplets is of no surprise.

Another tool that has been widely used for a modal degree assignment is an asymptotic period spacing. In sdBVs, in the asymptotic regime {\it i.e.} $n\gg l$, consecutive overtones of gravity modes are equally spaced in period \citep[e.g.][]{charpinet00,reed11}. Previous analysis of \kep\ photometric data of sdBVs showed that the average period spacing of dipole modes is nearly 250\,s \citep{reed18b}. The period spacings for higher degree modes can be calculated using the following relation $\Delta$P$_l$\,=\,P$_0$/$\sqrt{l(l+1)}$.

Multiplet helps to constrain the modal degree and provides a head start for determining the asymptotic period spacing, as such three peaks are assigned {\it l}\,=\,1 modes. In this work we could not rely on any multiplets and therefore we first focused on peaks that fit a sequence of {\it l}\,=\,1 radial overtones. Such peaks should be separated by around 250\,sec, not counting trapped modes. Peaks not satisfying our {\it l}\,=\,1 period spacing sequence were assumed to be {\it l}\,=\,2 modes or were unidentified {\it l} values if they have low amplitudes and do not fit the expected {\it l}\,=\,2 spacing. We found no convincing candidates for trapped {\it l}\,=\,1 modes. We stress though that without rotationally split modes, a modal degree assignment is not unique, since we try to fit an overtone sequence where some of the peaks may accidentally fit {\it e.g.} {\it l}\,=\,1, while being a quadrupole mode or a higher degree mode. Besides multiplets, another help would come from a pulsation spectrum that is dense enough to complete at least the {\it l}\,=\,1 sequence. We included the radial order and modal degree assignment in Tables\,\ref{tab:modeid1} -- \ref{tab:modeid3}. The average period spacings of dipole modes we estimated in these three stars are 247.22\,(65)\,s (\gaia\ DR2\,2129988841754560000), 208.70\,(56)\,s (\gaia\ DR2\,438585365733719168), 256.68\,(1.02)\,s (\gaia\ DR2\,537697439806045824). In the case of the first and third stars the average period spacing is close to the typical one found in sdBVs, while for the second star we found the spacing to be significantly lower. If this spacing is roughly correct it means that the core of \gaia\ DR2\,438585365733719168 is denser comparing to a typical sdB star, which suggests a more evolved phase of its EHB evolution, with central helium abundance Y$_{\rm c}$ < 0.1.

\begin{table}
\centering
\caption{List of frequencies detected in an amplitude spectrum of \gaia\ DR2\,2129988841754560000.}
\label{tab:modeid1}
\rowcolors{1}{}{light3}
\resizebox{\columnwidth}{!}{\begin{tabular}{ccccccc}
\hline
\rowcolor{light1}
& Frequency & Period & Amplitude & & & \\
\rowcolor{light1}
\multirow{-2}{*}{ID} &[$\upmu$Hz] & [s] & [ppt] & \multirow{-2}{*}{S/N} & \multirow{-2}{*}{\it l} &\multirow{-2}{*}{n} \\
\hline
f$_{\rm 1}$ & 66.327(7) & 15076.8(1.7) & 1.12(7) & 12.4 & 1 & 60 \\
f$_{\rm 2}$ & 72.865(18) & 13724.1(3.4) & 0.45(7) &  5.0 & - & - \\
f$_{\rm 3}$ & 76.6613(35) & 13044.4(6) & 2.29(7) & 25.4 & 1 & 52 \\
f$_{\rm 4}$ & 84.658(6) & 11812.2(9) & 1.29(7) & 14.3 & 1 & 47 \\
f$_{\rm 5}$ & 88.1774(13) & 11340.78(17) & 6.17(7) & 68.5 & 1 & 45 \\
f$_{\rm 6}$ & 92.286(6) & 10835.8(7) & 1.35(7) & 14.9 & 1 & 43 \\
f$_{\rm 7}$ & 94.737(11) & 10555.5(1.2) & 0.77(7) & 8.5 & 1 & 42 \\
f$_{\rm 8}$ & 101.685(18) & 9834.3(1.8) & 0.44(7) & 4.9 & 1 & 39 \\
f$_{\rm 9}$ & 109.870(15) & 9101.7(1.3) & 0.54(7) & 6.0 & 1 & 36 \\
f$_{\rm 10}$ & 118.077(18) & 8469.0(1.3) & 0.46(7) & 5.1 & - & - \\
f$_{\rm 11}$ & 118.9322(25) & 8408.15(18) & 3.29(7) & 36.5 & 2 & 59 \\
f$_{\rm 12}$ & 127.776(6) & 7826.19(36) & 1.38(7) & 15.4 & 1 & 31 \\
f$_{\rm 13}$ & 131.620(13) & 7597.6(7) & 0.65(7) & 7.2 & 1 & 30 \\
f$_{\rm 14}$ & 135.9947(38) & 7353.23(21) & 2.14(7) & 23.7 & 1 & 29 \\
f$_{\rm 15}$ & 144.718(17) & 6910.0(8) & 0.47(7) & 5.2 & - & - \\
f$_{\rm 16}$ & 148.195(16) & 6747.9(7) & 0.53(7) & 5.9 & - & - \\
f$_{\rm 17}$ & 149.1444(48) & 6704.91(21) & 1.72(7) & 19.1 & 2 & 47\\
f$_{\rm 18}$ & 150.457(6) & 6646.40(27) & 1.33(7) & 14.8 & 1 & 26 \\
f$_{\rm 19}$ & 161.9192(47) & 6175.92(18) & 1.73(7) & 19.2 & 1 & 24 \\
f$_{\rm 20}$ & 166.472(14) & 6007.0(5) & 0.58(7) & 6.4 & 2 & 42 \\
f$_{\rm 21}$ & 176.355(7) & 5670.37(24) & 1.10(7) & 12.2 & 1 & 22 \\
\hline
\end{tabular}}
\end{table}

\begin{table}
\centering
\caption{List of frequencies detected in an amplitude spectrum of \gaia\ DR2\,438585365733719168.}
\label{tab:modeid2}
\rowcolors{1}{}{light3}
\resizebox{\columnwidth}{!}{\begin{tabular}{ccccccc}
\hline
\rowcolor{light1}
& Frequency & Period & Amplitude & & & \\
\rowcolor{light1}
\multirow{-2}{*}{ID} &[$\upmu$Hz] & [s] & [ppt] & \multirow{-2}{*}{S/N} & \multirow{-2}{*}{\it l} &\multirow{-2}{*}{n} \\
\hline
f$_{\rm 1}$ & 20.449(45) & 48902(106) & 0.42(5) & 7.7 & - & - \\
f$_{\rm 2}$ & 71.436(46) & 13998.5(9.0) & 0.41(5) & 7.4 & 2 & 117 \\
f$_{\rm 3}$ & 91.073(26) & 10980.2(3.2) & 0.72(5) & 13.1 & 1 & 52 \\
f$_{\rm 4}$ & 94.650(21) & 10565.3(2.3) & 0.91(5) & 16.6 & 1 & 50 \\
f$_{\rm 5}$ & 98.520(17) &10150.2(1.7) & 1.14(5) & 20.7 & 1 & 48 \\
f$_{\rm 6}$ & 110.033(34) & 9088.2(2.8) & 0.61(6) & 11.2 & 1 & 43 \\
f$_{\rm 7}$ & 111.318(47) & 8983.3(3.8) & 0.45(6) & 8.2 & 2 & 75 \\
f$_{\rm 8}$ & 115.338(19) & 8670.2(1.5) & 0.98(5) & 17.8 & 1 & 41 \\
f$_{\rm 9}$ & 133.94(6) & 7466.2(3.1) & 0.34(5) & 6.2 & 1 & 35 \\
f$_{\rm 10}$ & 141.89(6) & 7047.5(2.8) & 0.34(5) & 6.1 & 1 & 33 \\
f$_{\rm 11}$ & 145.221(31) & 6886.1(1.5) & 0.62(5) & 11.2 & 2 & 57 \\
f$_{\rm 12}$ & 169.58(6) & 5897.0(2.1) & 0.31(5) & 5.6 & - & - \\
f$_{\rm 13}$ & 189.46(6) & 5278.2(1.6) & 0.34(5) & 6.1 & 2 & 43 \\
f$_{\rm 14}$ & 194.110(22) & 5151.7(6) & 0.88(5) & 16.0 & 1 & 24 \\
f$_{\rm 15}$ & 197.67(7) & 5058.8(1.7) & 0.29(5) & 5.3 & - & - \\
f$_{\rm 16}$ & 212.247(48) & 4711.5(1.1) & 0.40(5) & 7.2 & 1 & 22 \\
f$_{\rm 17}$ & 233.52(7) & 4282.3(1.2) & 0.28(5) & 5.0 & 1 & 20 \\
\hline
\end{tabular}}
\end{table}

\begin{table}
\centering
\caption{List of frequencies detected in an amplitude spectrum of \gaia\ DR2\,537697439806045824.}
\label{tab:modeid3}
\rowcolors{1}{}{light3}
\resizebox{\columnwidth}{!}{\begin{tabular}{ccccccc}
\hline
\rowcolor{light1}
& Frequency & Period & Amplitude & & & \\
\rowcolor{light1}
\multirow{-2}{*}{ID} &[$\upmu$Hz] & [s] & [ppt] & \multirow{-2}{*}{S/N} & \multirow{-2}{*}{\it l} &\multirow{-2}{*}{n} \\
\hline
f$_{\rm 1}$ & 123.167(8) & 8119.0(5) & 1.37(9) & 12.6 & 1 & 31 \\
f$_{\rm 2}$ & 141.626(7) & 7060.86(35) & 1.56(10) & 14.4 & 1 & 27 \\
f$_{\rm 3}$ & 154.153(22) & 6487.1(9) & 0.51(10) & 4.7 & 2 & 43 \\
f$_{\rm 4}$ & 165.5908(34) & 6038.98(13) & 3.18(9) & 29.4 & 1 & 23 \\
f$_{\rm 5}$ & 180.9512(29) & 5526.35(9) & 3.84(10) & 35.4 & 1 & 21 \\
f$_{\rm 6}$ & 195.113(20) & 5125.2(5) & 0.56(10) & 5.1 & 2 & 34 \\
f$_{\rm 7}$ & 274.501(13) & 3642.97(17) & 0.85(9) & 7.8 & 2 & 24 \\
f$_{\rm 8}$ & 313.988(11) & 3184.84(11) & 1.00(10) & 9.2 & 2 & 21\\
f$_{\rm 9}$ & 336.018(9) & 2976.03(8) & 1.21(10) & 11.2 & 1 & 11 \\
\hline
\end{tabular}}
\end{table}

\section{Orbital period stability}
\label{periods}
In total we found 33 eclipsing binary systems with sharp eclipses. Such eclipses are very suitable for deriving an orbital period and checking its stability that provides insights into physical effects in the systems, {\it e.g.} evolutionary changes (mass loss or mass exchange between components, tidal dissipation effects, gravitational radiation, magnetic braking), light travel time variation (stellar pulsations, a shift of the secondary eclipse, additional body in a system, apsidal motion). These effects alter the orbital period on a long time scale. In eclipsing binary systems orbital periods can be calculated from mid-times of primary or secondary eclipses. To estimate the mid-times we used the method described by \citet{kwee56}. Then, we fit the orbital periods to the mid-times and calculated the Observed-minus-Calculated (O-C) diagrams. The O-C diagram is a very effective tool to measure orbital period variation, if any, and a shape of a variation helps indicate its source. The interpretation of the shape is explained by \citet{sterken05}.

\begin{figure}
\includegraphics[width=0.50\textwidth]{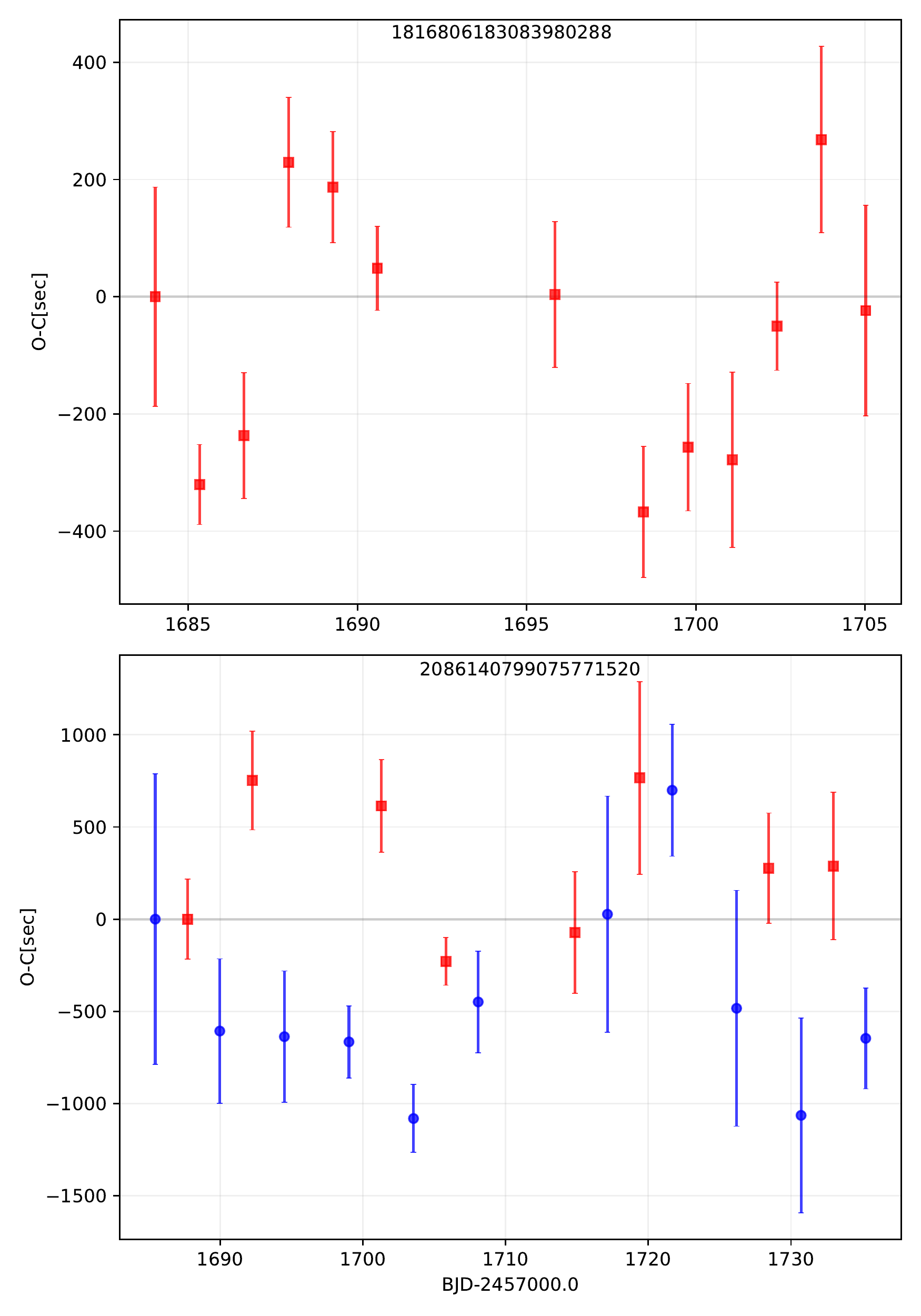}
\caption{Top panel shows the O-C diagram of Gaia\,DR2\,1816806183083980288. All eclipses are considered primary. Bottom panel shows the O-C analysis of Gaia\,DR2\,2086140799075771520. Mid-times of primary eclipses are plotted with red squares, while of secondary eclipses with blue circles.}
\label{fig:oc}
\end{figure}

Most of the selected eclipsing binaries in our sample show stable periods within the errors. It means that either the period does not change on a time scale comparable to the time coverage of our data or the change is too small to be detectable in the data. The typical time coverage of our data is only one month (one-sector). Out of 33 systems only in case of 12 the S/N ratio of the data is high enough to derive reasonable period estimation. We present the periods and the reference epochs, defined as a mid-time of one of the eclipses, in Table\,\ref{tab:periods}. In case of 10 systems the O-C diagrams show no variation within the errors, while in the case of two systems, \gaia\ DR2\,1816806183083980288 and \gaia\ DR2\,2086140799075771520, the variations look significant (Figure\,\ref{fig:oc}). We stress however that the errors of mid-times calculated using the method reported by \citet{kwee56} are underestimated, hence the true ones can be larger making the variations insignificant. The shapes and the physical reasons of these variations, if real, are not clear and require longer coverage to make any conclusive implications.

\begin{table}
\centering
\caption{Ephemerides of a selected sample of eclipsing binaries, calculated from mid-times of primary eclipses.}
\label{tab:periods}
\rowcolors{1}{}{light3}
\resizebox{\columnwidth}{!}{\begin{tabular}{lll}
\hline
\rowcolor{light1}
& \multicolumn{1}{c}{Period} & \multicolumn{1}{c}{T$_0$} \\
\rowcolor{light1}
\multicolumn{1}{c}{\multirow{-2}{*}{\gaia\,DR2}} & \multicolumn{1}{c}{[days]} & \multicolumn{1}{c}{[days]} \\
\hline
2129388572827118720 & 1.23135725(49)& 2458684.75482(12)\\
190164354254168192 & 2.34580(14) & 2458817.8620(8)\\
1816806183083980288 & 1.31229(17) & 2458684.0236(6)\\
1846629538332584960 & 0.8097(1) & 2458711.2610(9) \\
2093326416800046848 & 4.1776590(9) & 2458684.29845(44)\\
450362269138662400 & 3.4574(9) & 2458792.6052(27)\\
967312053368970624 & 3.185675(7) & 2458843.19265(35)\\
1831073927011326720 & 7.47899(39) & 2458681.13135(50)\\
2086140799075771520 & 4.5233(5) & 2458683.2089(31)\\
1821225425253990400 & 1.177035(9) & 2458683.7224(12)\\
431584053656962816 & 6.3148495(40) & 2458765.19978(15)\\
4322472232203849856 & 3.4800(6) & 2458684.5481(18)\\
\hline
\end{tabular}}
\end{table}

\section{Summary}
\label{sec:summary}
In Paper I we reported our result of the search for sdBVs in the southern ecliptic hemisphere observed by \tess\ satellite and we detected two convincing cases, while the remaining 26 sdBs turned out to be variable, and in most cases their binary nature causes the variations. The total number of variables found was 1807. In this work, we have continued our effort to search for sdBVs across the sky using \tess\ satellite. In total, we found 506 variables, among which 13 are likely sdBV stars, while another 40 sdBs show other types of variability, most from being in a binary system. One of them shows \hwvir\ shape of the phased time-series data. We found another five objects showing \hwvir\ variations among spectroscopically unclassified, bringing the final number of new \hwvir\ systems to seven (if all of them are confirmed to be sds). Comparing the number of new sdB variable objects, we can notice that even though the number of all objects, available in \citet{geier20}, and consequently the total number of new variables, was larger in the southern ecliptic hemisphere, we found more variable sdBs, including sdBVs, in the northern ecliptic hemisphere. Since, we rejected objects observed in the short cadence mode, we cannot conclude on the larger number of them in the north, but it clearly means there was still more to find.

Summing up all variable stars we found in the \tess\ field of view, we arrive with the following numbers: 15 likely sdBVs, 66 variable non-pulsating sdBs, 33 variable sds, 2076 spectroscopically unclassified objects (including 113 objects that show peaks in the sdBV g-mode region), 123 non-sdB variables. The ultimate goal of our work was to discover missing sdBVs to contribute to the completeness of the sdBV sample in the \tess\ mission. The byproduct of our findings is additional objects for the short cadence observations (either 2\,min or 20\,sec) during the second run of TESS in the southern and northern ecliptic hemispheres. The extra runs can deliver more precise data for further astrophysical analysis. Another byproduct will be pushing the Nyquist frequency beyond the p-mode region in sdBVs. For now, our search is limited to g-mode sdBVs, which is a consequence of the 30\,min. cadence, however the search utilizes the most updated sdB database along with the all-sky space survey, allowing for the most complete sample of g-mode sdBVs currently possible. Spectroscopic classification is needed to confirm 113 objects that show sdBV-like amplitude spectra, as well as all sds variables and the massive number of other unclassified variables.

The sdBVs we found in this work are not rich in g-modes that is why are not best cases for any mode identification. However, likewise in Paper I, among objects that are not confirmed sdBs but show typical g-mode sdBV amplitude spectra, we have selected three objects best-suited for the mode identification. They show the richest amplitude spectra. We searched for multiplets and found none, therefore we used only period spacing for the modal degree assignment. The sequences of presumably same degree overtones are not too complete but the multiples of 250\,sec (ish) was still found. This may be another argument for these objects being sdB stars. One of the stars shows 208\,sec spacing which is either an indication of a denser core or incorrect assumption that the star is an sdB. Our identification will only be reliable if these objects are spectroscopically confirmed to be sdBs.

We selected a sample of 12 eclipsing binary systems with sharp eclipses and derived mid-times of eclipses in order to calculate O-C diagrams. We measured orbital periods and checked their stabilities. Out of 12 systems only two show significant O-C variation, within the given errors, suggesting that some period change may be present in those systems. This conclusion is uncertain as the errors are underestimated, and longer data coverage is needed to make any physical interpretations.

\section*{Acknowledgements}
Financial support from the Polish National Science Center under projects No.\,UMO-2017/26/E/ST9/00703 and UMO-2017/25/B/ST9/02218 is acknowledged. This paper includes data collected by the \tess\ mission. Funding for the \tess\ mission is provided by the NASA Explorer Program. This work has made use of data from the European Space Agency (ESA) mission \gaia\ (\url{https://www.cosmos.esa.int/gaia}), processed by the \gaia\ Data Processing and Analysis Consortium (DPAC, \url{https://www.cosmos.esa.int/web/gaia/dpac/consortium}). Funding for the DPAC has been provided by national institutions, in particular the institutions participating in the \gaia\ Multilateral Agreement. Fruitful remarks from an anonymous referee are appreciated.

\section*{Data availability}
The datasets were derived from MAST in the public domain archive.stsci.edu.




\bibliographystyle{mnras}
\bibliography{myrefs} 








\bsp	
\label{lastpage}
\end{document}